\documentclass[a4paper,11pt]{article}
\usepackage{graphicx,amssymb,amsmath,bm,latexsym,color,epsf,cite,verbatim, mathtools}

\usepackage{hyperref}

\makeatletter
\@addtoreset{equation}{section}

\makeatother
\newcommand{\be}{\begin{equation}}
\newcommand{\ee}{\end{equation}}
\newcommand{\bea}{\begin{eqnarray}}
\newcommand{\eea}{\end{eqnarray}}
\newcommand{\vs}[1]{\vspace{#1 mm}}

\newcommand{\Tr}{{\rm Tr}}

\def\eff{\mathbf{f}}
\def\ess{\mathbf{S}}
\def\ti{\mathbf{T}}
\def\cut{C}

\begin{document}

\begin{center}
{\Large\bf The cosmological constant problem\vspace{2mm}
\\ and the effective potential of a gravity-coupled scalar
%\\
%or
%\\
%Gravity regulates the scalar effective potential
}

\vs{8}

{\large
Renata Ferrero\footnote{e-mail address: renata.ferrero@fau.de}
and Roberto Percacci\footnote{e-mail address: percacci@sissa.it}$^{,3}$
} \\
\vs{8}
$^1${\em Institute for Quantum Gravity, Friedrich-Alexander-Universität Erlangen-Nürnberg, Staudtstr. 7, 91058 Erlangen, Germany}

$^2${\em \textit{International School for Advanced Studies, via Bonomea 265, 34136 Trieste, Italy}}

$^3${\em \textit{INFN, Sezione di Trieste, Italy}}

\end{center}
\vs{1}

\setcounter{footnote}{0} 
%%%%%%%%%%%%%%%%%%%%%%%%%%%%%%%

\begin{abstract}
We consider a quantum scalar field in a classical (Euclidean) De Sitter background,
whose radius is fixed dynamically by Einstein's equations.
In the case of a free scalar, it has been shown by Becker and Reuter
that if one regulates the quantum effective action 
by putting a cutoff $N$ on the modes of the quantum field,
the radius is driven dynamically to infinity when $N$ tends to  infinity.
We show that this result holds also in the case of a self-interacting scalar,
both in the symmetric and broken-symmetry phase.
Furthermore, when the gravitational background is put on shell,
the quantum corrections to the mass and quartic self-coupling
are found to be finite.
\end{abstract}

\normalsize
%%%%%%%%%%%%%%%%%%%%%%%
\section{Introduction}

The motivation for this work came originally from the cosmological constant problem.
A common way of presenting this problem proceeds along the following lines
(see e.g. \cite{Weinberg:1988cp}).
By Lorentz invariance, the vacuum expectation value (VEV) of the energy-momentum tensor of a quantum field should be of the form
\footnote{Here we work in signature $-+++$, but the rest of the paper is based
on Euclidean calculations.}
\be
\langle T_{\mu\nu}\rangle_{vac}=-\langle\rho\rangle_{vac}\,g_{\mu\nu}\ ,
\label{Tvac}
\ee
and thinking of the field as an infinite collection of oscillators,
the vacuum energy density is the sum of the vacuum energies of all the oscillators.
In flat space, putting a cutoff $C$ on the spatial momenta,
\footnote{We reserve the symbol $\Lambda$ for the cosmological constant.} 
one gets
\be
\langle\rho\rangle_{vac}
=\int_0^C\frac{d^3q}{(2\pi)^3}\frac12\sqrt{q^2+m^2}
\sim\frac{\cut^4}{(4\pi)^2}\ .
\label{pauli}
\ee
The same conclusion can be reached by calculating the Euclidean effective action of
a quantum field in a background metric $g_{\mu\nu}$:
there is a quartically divergent term proportional to the spacetime volume.
(This calculation is reviewed in Appendix A).
Using this in the Einstein equations leads to a curvature that
grows like a power of the cutoff.
The first argument of this type is generally credited to Pauli (unpublished, see \cite{straumann}),
who found that if the cutoff $C$ is of the order of the mass of the electron,
the universe would not extend further than the Moon.
Nowadays we know that quantum field theory is valid way beyond
the scale of the electron, and the problem becomes much worse 
\cite{Zeldovich:1967gd}.
One could also assume that a ``bare'' cosmological term
cancels most or all of the vacuum energy, but this is ad hoc
and leaves us with an enormous fine tuning problem.

The calculation in (\ref{pauli}) has been criticized 
for various reasons by several authors
\cite{Akhmedov:2002ts,Ossola:2003ku,Maggiore:2010wr,Asorey:2012xq}
\footnote{see also \cite{DeWitt:1975ys}.}
The criticism of Akhmedov \cite{Akhmedov:2002ts}, and then by Ossola and Sirlin \cite{Ossola:2003ku}, was that the momentum cutoff breaks
the Lorentz invariance that was assumed in (\ref{Tvac}).
By considering alternative regularization methods that preserve Lorentz invariance,
they concluded that the vacuum energy of free quantum fields 
cannot diverge more than quadratically.
This is significant, but not enough to solve the problem,
and in any case interacting fields could still give rise to a quartic divergence.

A more radical critique has appeared recently
in a paper of Becker and Reuter \cite{Becker:2020mjl}.
They claim that the estimate (\ref{pauli}) is not self-consistent,
in the sense that the VEV of the energy-momentum tensor is
calculated in one metric (usually the Minkowski metric) and then used in another one.
A self-consistent calculation would amount to calculating the
VEV of the energy-momentum tensor in the same metric that
solves the Einstein equations.
The general argument for using the flat space estimate (\ref{pauli}) in curved spacetime,
is that the divergences of quantum fields are universal,
because every manifold is flat on very small scales.
Becker and Reuter argue that this is misleading
and does not reflect correctly the physics.
Instead, they show that a background-independent calculation leads
to opposite conclusions, namely the curvature of the metric
{\it decreases} as more modes of the quantum field are taken into account,
and goes to zero in the limit when the cutoff goes to infinity.
The calculation is done on a Euclidean De Sitter space (a sphere)
and there are two main ways in which it deviates from the standard one. 
The first is the use of a dimensionless cutoff,
which is especially natural on a compact manifold, where the spectrum 
of the Laplacian is discrete.
The second is that one has to take into account the backreaction of the quantum field {\it before} 
sending the cutoff to infinity.
We will review this calculation for a free scalar field in Section 2.

The original calculation for a free scalar on $S^4$ has been later generalized
to gravitons of $S^4$ \cite{Becker:2021pwo}
and a scalar on a hyperboloid \cite{Banerjee:2023ztr}.
The latter calculation shows that the use of dimensionless cutoffs
can be extended also to the case when the spectrum is continuous.

The main aim of this work is to extend the Becker-Reuter results
for the cosmological constant problem to the self-interacting case
and to evaluate the main properties of the effective action.
In contrast to the free case, in a self-interacting theory the mass receives quantum corrections,
that in standard QFT calculations are quadratically divergent
and pose fine tuning problems that are similar to those of the cosmological constant.
One motivation of this work was the hope that the procedure 
that removes the fine tuning of the cosmological constant
may also be able to alleviate the fine tuning of the mass.
Another motivation came from arguments of observability.
What do we mean when we say that a universe containing $N$ modes 
of a single massless scalar field has a certain curvature,
and that the curvature decreases with $N$?
One seems to be implicitly admitting the existence of an observer
outside the universe that can compare its scale to that of another universe
with a different number of modes $N'$.
It would be better to make ``relational'' statements,
for example by saying that the ratio $R/m^2$ has a certain value,
where $m$ is the mass of some field.
\footnote{This is in practice what we do when we say that our universe is large
compared to a hydrogen atom, for example.}
Since this mass receives quantum corrections,
it could conceivably happen that $m^2$ scales with $N$ in the same way as $R$,
in which case the physical meaning of the Becker-Reuter results
would be much weakened.

The outcome of our calculations is encouraging.
First of all, we find that the behavior of the curvature is not modified
by the presence of the interactions, while the VEV of the scalar (when nonzero)
has a finite limit when the cutoff goes to infinity.
The decreasing behavior of curvature as a function of the cutoff thus appears
to be a robust feature. 
What is more surprising, we find that when the metric is put on shell,
the quantum corrections to the mass are finite.
Both the quadratic and the logarithmic corrections that arise in the standard
way of calculating, are canceled.
This removes the fine tuning problems and also
turns the Becker-Reuter result into a statement about the
dimensionless ratio $R/m^2$, which is arguably a more physical quantity
than $R$ or $m^2$ separately.
In addition, we find that also the quantum corrections to the
scalar self-coupling, that usually are logarithmically divergent, are finite
when the metric is put on shell.

We emphasize that all our results are about semiclassical gravity,
also including the effect of the backreaction.
We do not make any statements about quantum gravity.

The plan of the paper is as follows.
In Section 2 we review the arguments of Becker and Reuter
concerning the cosmological constant problem for a free scalar field,
possibly also in the presence of a nonminimal coupling to curvature.
Section 3 contains general formulae for the case of the self-interacting scalar.
Section 4 then gives our main results for the symmetric phase,
while in Section 5 we consider the broken symmetry phase.
A discussion of the results is given in Section 6.
Appendix A contains, for the sake of comparison,
the calculation of the effective action with heat kernel methods
and a dimensionful cutoff.
Appendix B is a summary of properties of certain special functions.

%%%%%%%%%%%%
\section{Free scalar field}
%%%%%%%%%%%%

In this section we review and extend the results of \cite{Becker:2020mjl}.
We assume that we are in a semiclassical regime where the metric $g_{\mu\nu}$ 
can be treated as a classical field with the (Euclidean) Hilbert action (and cosmological term)
\be
S_H(g)=\frac{1}{16\pi G}\int d^4x\sqrt{g}\left[2\Lambda-R\right]\ ,
\ee
interacting with a quantum scalar field $\phi$ with action
\be
S_m(\phi;g)=\int d^4x\sqrt{g}\left[\frac12(\partial\phi)^2
+\frac12\xi R\phi^2
+\frac12 m^2\phi^2+\frac{1}{4!}\lambda\phi^4\right]\ .
\ee
The backreaction of the scalar field on the metric is encoded in the effective action (EA).
At one loop it is given by the familiar formula
\be
\Gamma(g,\phi)=S_H(g)+S_m(g,\phi)+\frac12\Tr\log(\Delta/\mu^2)
\label{effac}
\ee
where 
\be
\Delta=-\nabla^2+E\ ,\qquad
E=\xi R+m^2+\frac12\lambda\phi^2
\label{oper}
\ee
and $\mu$ is a suitable scale that does not appear in the equations of motion
(for example one could set $\mu=m$).
In this section we restrict ourselves to a free scalar field, thus we set $\lambda=0$ and $m^2>0$.

Variation of the EA with respect to the metric yields the
semiclassical Einstein equations
\be
R_{\mu\nu}-\frac12 R g_{\mu\nu}+\Lambda_B g_{\mu\nu}
=8\pi G \langle T_{\mu\nu}\rangle
\label{eeq}
\ee
where the l.h.s. comes from the classical Hilbert action (with a bare cosmological term)
and the r.h.s is the VEV of the energy-momentum tensor of the quantum field.

%%%%%%%%%%%%%%%%%%%%%%%%%
\subsection{Spectrum, cutoff and effective action on $S^4$}
\label{becker_reuter}

For our purposes it is enough to study the EA on a Euclidean De Sitter space,
i.e. a sphere $S^4$.
The metric of the sphere is almost completely determined by $O(5)$ symmetry,
the only remaining degree of freedom being the radius $r$, or equivalently
the (constant) scalar curvature $R=12/r^2$.
In the following we shall therefore obtain the equation of motion
by simply deriving the action with respect to $R$.
For example, recalling that the volume of the four-sphere is
\be
V_4=\frac{384\pi^2}{R^2}\ ,
\ee
the Hilbert action for a spherical metric can be written
\be
S_H(R)=\frac{48\pi\Lambda_B}{GR^2}-\frac{24\pi}{GR}
\ee
and deriving with respect to $R$ we obtain the familiar equation
\be
R=4\Lambda\ .
\ee

One great advantage of working on a sphere is that the spectrum of the Laplacian is well-known
and therefore it is possible to calculate the EA without resorting
to heat kernel asymptotics.
Furthermore, it makes the use of a dimensionless cutoff particularly natural.
The Laplacian $-\nabla^2$ on $S^4$ has eigenvalues $\lambda_\ell$ with multiplicity $m_\ell$,
given by
\be
\lambda_\ell=\frac{R}{12}\ell(\ell+3)\ ,\quad
m_\ell=\frac16 (\ell+1)(\ell+2)(2\ell+3)\ ,\quad
\ell=1,2\ldots
\label{spectrum}
\ee
The operation $\Tr$ in the definition (\ref{effac}) of the EA
is a functional trace that on the sphere can be written explicitly as a sum
over all eigenstates of the Laplacian.
We regulate the sum by putting an upper bound $N$ on the quantum number $\ell$:
\be
\frac12\Tr_N\log(\Delta/\mu^2)=\frac12\sum_{\ell=1}^N m_\ell\log\left(\frac{\lambda_\ell+E}{\mu^2}\right)\ .
\ee
We should contrast this to the more standard choice of cutting off the
sum at some cutoff $C$ with dimension of mass, such that
\be
\lambda_\ell<C^2\ .
\ee
At this stage there is no significant difference between the two procedures,
because the dimensionless and the dimensionful cutoffs are simply related by
\be
C^2=\frac{R}{12}N(N+3)\ .
\label{ddcut}
\ee
However, we shall see in the following that when we demand that the background metric
satisfy Einstein's equations, the two procedures lead to very different conclusions.

With the $N$-cutoff in place, the total number of modes that is included in the trace is
\be
\eff(N)=\sum_{\ell=1}^N m_\ell
=\frac{1}{12}N(N+4)(N^2+4N+7)\ .
\label{eff}
\ee

%%%%%%%%%%%%%%%%%%%%%%
\subsection{Massless free field}

If we put $m=\lambda=\xi=0$ we have a massless free scalar field with vanishing expectation value
and kinetic operator $\Delta=-\nabla^2$.
Its classical action vanishes and the EA depends only on the metric:
\bea
\Gamma_N(R)&=&S_H(R)+\frac12\Tr_N\log(-\nabla^2/\mu^2)
\nonumber\\
&=&
V_4\left[\frac{\Lambda_B}{8\pi G}
-\frac{1}{16\pi G}R\right]
+\frac12\sum_{\ell=1}^N m_\ell \log\left(\frac{R}{12\mu^2}\ell(\ell+3)\right)
\nonumber\\
&=&\frac{48\pi\Lambda_B}{GR^2}
-\frac{24\pi}{GR}
+\frac12\eff(N) \log\left(\frac{R}{12\mu^2}\right)
+\frac12\ti(N,0)\ ,
\label{gammaN}
\eea
where we split the log into a $R$-dependent and an $R$-independent term.
We defined the dimensionless function
\be
\ti(N,z)=\sum_{\ell=1}^N m_\ell \log\left(\ell(\ell+3)+z\right)\ ,
\ee
representing the quantum trace for a sphere of radius $R=12\mu^2$.
In this first application, $z=0$ and $\ti(N,0)$ is just a field-independent,
quartically divergent constant.
Notice that the last two terms in (\ref{gammaN}), that are of quantum origin,
do not renormalize the classical gravitational couplings.
In particular, there is no quartically divergent renormalization of the cosmological constant.
\footnote{This is somewhat reminiscent of unimodular gravity,
where the quartically divergent term is independent of the metric.
Note that the last term can be interpreted as $\int d^4x \sqrt{g}R^2$,
which is scale-independent.}

\begin{figure}[]
\centering
\includegraphics[width=.6\textwidth]{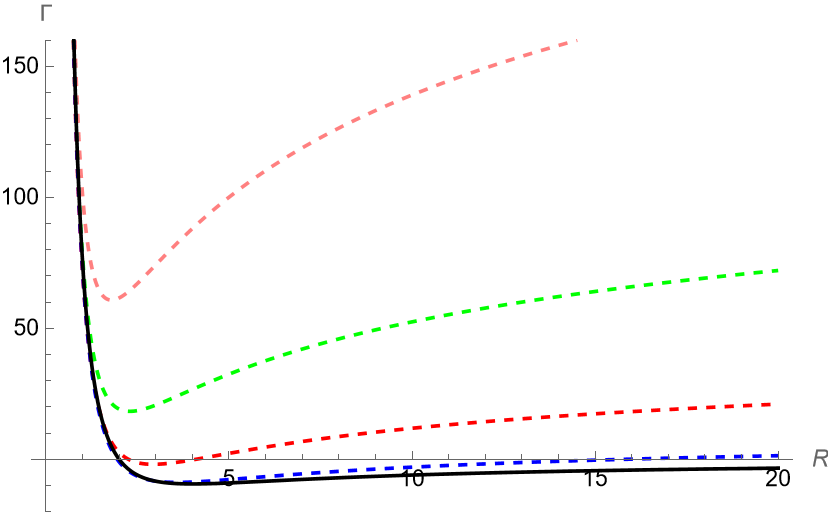}
\caption{The EA as a function of $R$ for 
a free massless scalar and $G=\Lambda=\mu=1$
and increasing $N$. The continuous black curve is the
classical action. It has a minimum at $R=4$ 
and asymptotes to zero for $R\to\infty$.
The dashed curves correspond to adding the first 1, 2, 3 or 4 modes.
The minimum moves up and to the left for increasing $N$,
and the second derivative at the minimum increases.}
\label{fig:gamma}
\end{figure}

Both terms of the classical action go to zero for large $R$,
so the logarithmically growing quantum term dominates in this regime.
On the other hand, for $R\to 0$ the cosmological term dominates and diverges to $\pm\infty$,
depending on its sign.
Thus, in the presence of a positive bare cosmological constant
the EA must have a minimum as a function of $R$.
(See Figure \ref{fig:gamma}.)
To find it we derive the EA with respect to $R$, arriving at the equation
\be
\frac{24\pi}{GR^3}(-R+4\Lambda)
=\frac{1}{2R}\eff(N)\ ,
\label{eommassless}
\ee
The solution of this equation is a sphere of curvature
\bea
R&=&\frac{24\pi}{G\eff(N)}\left(-1\pm\sqrt{1+\frac{G\Lambda\eff(N)}{3\pi}}\right)
\nonumber\\
&\approx& 
%8\sqrt{3\pi}\sqrt{\frac{\Lambda_B}{G}}\left(\frac{1}{\sqrt{\eff(N)}}+O(1/\eff(N))\right)
%\nonumber\\
%&\approx&
48\sqrt{\frac{\pi\Lambda}{G}}
\left(
\frac{1}{N^2}-\frac{4}{N^3}\right)
+\left(\frac{600\sqrt{\pi\Lambda}}{\sqrt G}-\frac{288\pi}{G}\right)\frac{1}{N^4}
+\left(-\frac{1728\sqrt{\pi\Lambda}}{\sqrt G}+\frac{2304\pi}{G}\right)\frac{1}{N^5}
\nonumber\\
&&
+\left(\frac{5106\sqrt{\pi\Lambda}}{\sqrt G}+\frac{864\pi^{3/2}}{G^{3/2}\sqrt\Lambda}-\frac{11808\pi}{G}\right)\frac{1}{N^6}
+\left(-\frac{16728\sqrt{\pi\Lambda}}{\sqrt G}-\frac{10368\pi^{3/2}}{G^{3/2}\sqrt\Lambda}+\frac{49536\pi}{G}\right)\frac{1}{N^7}
\nonumber\\
&&
+O(1/N^8)
\ ,
\label{silvia}
\eea
where $\approx$ means that we take the dominant behavior for large $N$
and in the expansion we have selected the solution with the upper sign.
This is Becker and Reuter's main result.
We see that, opposite to standard lore,
the curvature of spacetime {\it decreases}
when more quantum modes are included in the calculation.

Note that whereas the size of the universe depends on $\Lambda$,
one can always make the universe as large as one wants by choosing $N$
sufficiently large.
Thus, there is no reason to think of $\Lambda$ as being small.
In fact, in the following, we will always assume that $\Lambda$
is of order one in Planck units.

It is also worth noting that the quantum term in the EA,
that gives rise to this behavior, is the one that generates the trace anomaly.
This should not come as a surprise, given that the only dynamical degree of freedom
of the metric is its overall scale.
In fact, on a sphere the only independent component of Einstein's equations (\ref{eeq}) is the trace
\be
-R +4\Lambda_B 
=8\pi G\,\frac{ \int d^4x\sqrt{g}\langle T^\mu{}_\mu\rangle}{\int d^4x\sqrt{g}}\ ,
\label{eom}
\ee
where in the r.h.s. we exploited homogeneity.
Comparing this to (\ref{eommassless}) and recalling that
the classical system was scale invariant, we find that
\be
\int d^4x\sqrt{g}\langle T^\mu{}_\mu\rangle=\eff(N)
\ee
is the trace anomaly.

In \cite{Becker:2020mjl} an alternative calculation has been given
where a direct definition of the VEV of the energy-momentum tensor,
based on the quantization of the classical energy-momentum tensor, gives
\be
\int d^4x\sqrt{g}\langle T^\mu{}_\mu\rangle=-\eff(N)\ .
\ee
With this definition, the semiclassical Einstein equations have
a solution even when the bare cosmological constant is zero.
It is remarkable that in spite of the different sign, both definitions
ultimately lead to similar conclusions.
In this paper we shall stick to the first definition,
where $\langle T^\mu{}_\mu\rangle$ is defined
by a variational procedure applied to the quantum EA.
\bigskip

%%%%%%%%%%%%%%%%%%%%%%%%%%
\subsection{Nonminimal coupling}
%%%%%%%%%%%%%%%%%%%%%%%%%%

\begin{figure}[]
\centering
\includegraphics[width=.48\textwidth]{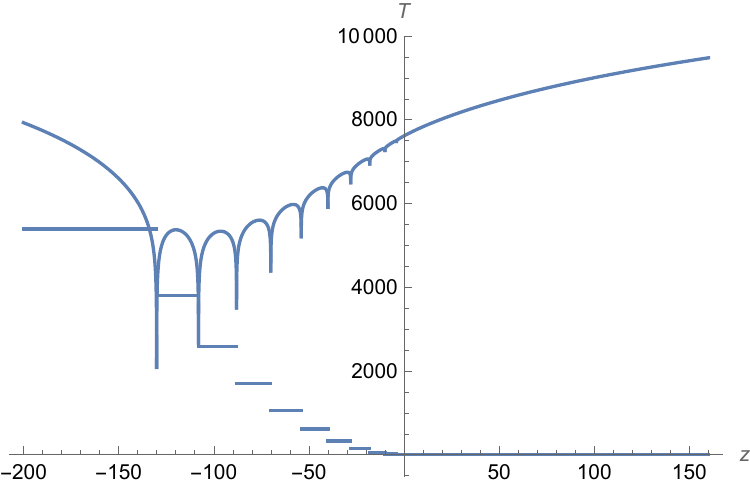}\quad
\includegraphics[width=.48\textwidth]{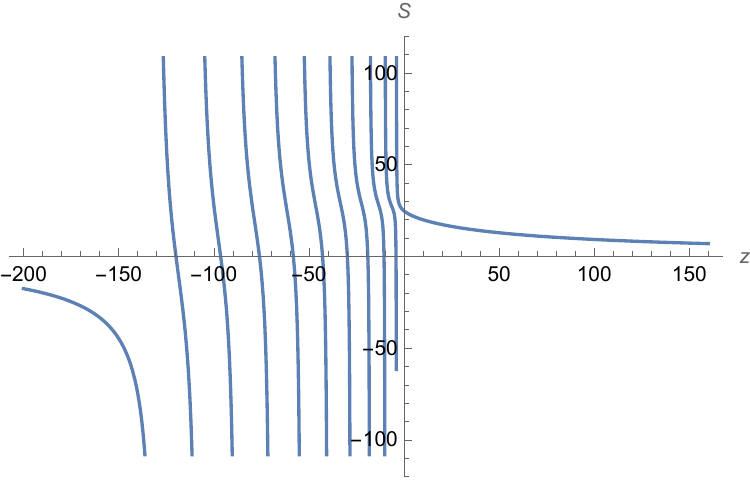}
\caption{Left panel: the real and imaginary parts of the function $\ti(10,z)$.
Right panel: the  function $\ess(10,z)$ defined in (\ref{ess}).}
\label{fig:ess}
\end{figure}

Before coming to the massive case, we consider briefly the effect
of a nonminimal coupling $\tfrac12\xi \phi^2 R$.
For a constant scalar, the equation of motion implies $\phi=0$.
We thus do not need to consider other backgrounds.
The EA (\ref{effac}) on a spherical background is now:
\bea
\Gamma_N(R)&=&S_H(R)+\frac12\Tr_N\log\left(\frac{-\nabla^2+\xi R}{\mu^2}\right)
\nonumber\\
&=&V_4\left[\frac{\Lambda_B}{8\pi G}
-\frac{1}{16\pi G}R
%+\frac12\xi\phi^2 R
\right]
+\frac12\sum_{\ell=1}^N m_\ell \log\left(\frac{R}{12\mu^2}(\ell^2+3\ell+12\xi)\right)
\nonumber\\
&=&\frac{48\pi\Lambda_B}{GR^2}
-\frac{24\pi}{GR}
%+\frac{192\pi^2\xi\phi^2}{R}
+\frac12\eff(N) \log\left(\frac{R}{12\mu^2}\right)
+\frac12\ti(N,12\xi)\ .
\eea
We note that $\ti(N,12\xi)$ is again a field-independent constant,
so this term does not affect the equations of motion.
However, now that the second argument of $\ti$ is nonzero,
there are additional issues that may arise.
Figure \ref{fig:ess} shows the function $\ti(10,z)$ and its derivative $\ess(10,z)$.
They are seen to have poles at $z=-\ell(\ell+3)$ for $\ell=1,\ldots,10$,
that correspond to points where the arguments of the logs in the sum become zero.
Furthermore $\ti(N,z)$ becomes complex for $z<-4$.
In order to avoid this, we assume $\xi>-1/3$.
The gravitational equation of motion is again (\ref{eommassless}),
so the solutions of the simultaneous equations of motion
are identical to those of the case $\xi=0$ discussed previously.

%%%%%%%%%%%%%%%
\subsection{The massive case}
%%%%%%%%%%%%%%%

We gradually increase the complication by adding a mass term.
Once again we can set $\phi=0$ by its equation of motion.
Then the EA for the metric becomes
\bea
\Gamma_N(R)\!\!\!&=&\!\!\!S_H(R)
+\frac12\Tr_N\log\left(\frac{-\nabla^2+m^2+\xi R}{\mu^2}\right)
\nonumber\\
&=&\!\!\!V_4\left[\frac{\Lambda_B}{8\pi G}
-\frac{1}{16\pi G}R
%+\frac12m^2\phi^2+\frac12\xi\phi^2 R
\right]\!
+\frac12\sum_{\ell=1}^N m_\ell \log\left(\!\frac{R}{12\mu^2}(\ell^2+3\ell+12\xi)+\frac{m^2}{\mu^2}\right)
\nonumber\\
&=&\!\!\!\frac{48\pi\Lambda_B}{GR^2}
-\frac{24\pi}{GR}
%+\frac{192\pi^2(m^2+\xi R)\phi^2}{R^2}
+\frac12\eff(N) \log\left(\frac{R}{12\mu^2}\right)
+\frac12\ti\left(N,\frac{12m^2}{R}+12\xi\right)\ .
\label{gammaNxi}
\eea
The quantum trace is independent of $\phi$, so the scalar potential
does not receive any quantum corrections.
Since the sum is finite, we can take the derivative under the sum,
so that the equation of motion for the metric can be written in the form
\be
\frac{24\pi}{GR^3}(-R+4\Lambda)
=\frac{1}{2R}\eff(N)
-\frac{6m^2}{R^2}\ess\left(N,\frac{12m^2}{R}+12\xi\right)\ ,
\label{cadmo}
\ee
where we introduced a new dimensionless function
\be
\ess(N,z)\equiv\frac{\partial\ti(N,z)}{\partial z}
=\sum_{\ell=1}^N
m_\ell\frac{1}{\ell(\ell+3)+z}\ .
\label{ess}
\ee
An explicit formula for this function in terms of harmonic numbers is given
in (\ref{essexp}), but we see already from this definition that
this function diverges quadratically with $N$, for fixed $z$.
It is clear that in the limit $m^2\to 0$ and $\xi\to 0$ at fixed $R$ we get back (\ref{eommassless}).
For $z\ll1$ we have just a small deformation of the massless solution.
We are interested in the physically more realistic case $z\gg 1$,
when, in the corresponding  Lorentzian world,
the Compton wavelength of the scalar particles
is much smaller than the Hubble radius.

\begin{figure}[]
\centering
\includegraphics[width=.6\textwidth]{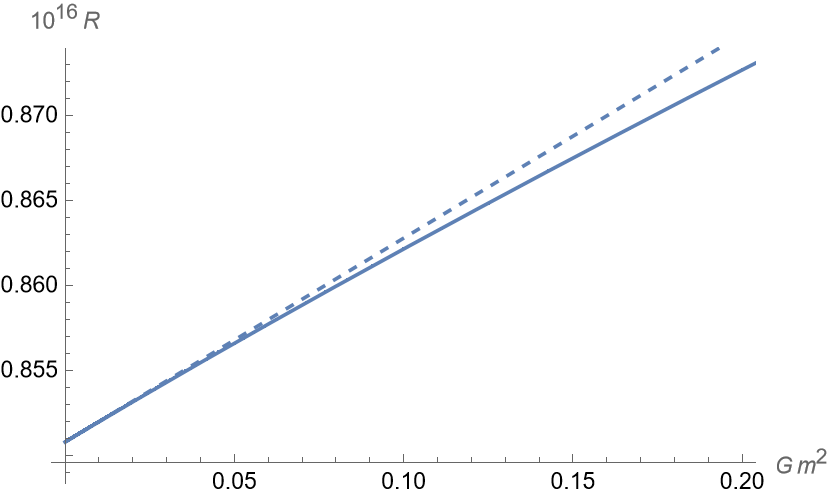}
\caption{The curvature $R$ as a function of $m^2 G$, for $G\Lambda=1$, $\xi=0$
and $N=10^9$.
Continuous line: numerical solution of (\ref{cadmo}),
dashed line: the iterative solution, to linear order in $m^2$,
which is practically indistinguishable from its first iteration (\ref{firstiteration}).}
\label{fig:massive}
\end{figure}

When $m^2$ is not zero, $R$ appears explicitly in the function $\ess$,
and this makes the equation of motion unsolvable analytically.
Motivated by the result of the massless case, let us make the ansatz
\be
R=\frac{K_2}{N^2}+\frac{K_3}{N^3}+\frac{K_4}{N^4}+\ldots
\label{gloria}
\ee
and check whether it is consistent.
For the leading behaviour we can keep only the first term with coefficient $K_2$.
In this case the second argument of $\ess$ is
\be
z=yN^2+\ldots\ ,
\label{zy}
\ee 
where $y=\tfrac{12m^2}{K_2}$.
Note that the $\xi$-term is negligible in this approximation.
We insert the ansatz in the equation of motion and use the expansion
(\ref{Sexp}) of $\ess\left(N,yN^2\right)$ for large  $N$.
When one makes the ansatz for $R$, the equation of motion becomes 
a divergent series in $N$ with highest power $N^6$, and the logarithms are finite.
\footnote{We refer here to the equation of motion in the form (\ref{cadmo}).
If $R$ is assumed nonzero, we can remove factors of $R$ from the denominators,
and this lowers the degree of the divergence.
If we remove the volume factor, the leading divergence is $N^2$.
If we write the equations in
the standard way $R=4\Lambda+\ldots$, there are no divergences at all,
due to the already noticed automatic cancellation of the $\log N$ terms.}
The coefficient of the highest divergence is
\be
\frac{1}{24K_2^3}\left[
K_2^2-\frac{2304\pi\Lambda}{G}-24K_2 m^2
+288m^4\log\left(1+\frac{K_2}{12m^2}\right)
\right]\ .
\ee
We can set it to zero by fixing $K_2$.
Due to the presence of $K_2$ in the log, this can only be done numerically,
after $m^2$ has been fixed.
An example of a numerical solution, as a function of $m^2 G$
and for  specific values of $\xi$ and $\Lambda G$ is given in Figure \ref{fig:massive}.
In order to have an analytic formula we can expand to first order in $m^2$,
leading to a second order algebraic equation whose solution is (\ref{firstiteration}) below.

The other coefficients in (\ref{gloria}) can be determined iteratively.
We can remove all the divergences from the equation of motion
by expanding $R$ up to order $N^{-8}$.
We insert the ansatz in the equation of motion and expand it up to order $N^0$.
We thus arrive at an equation of the form
$$
\frac{C_6}{K_2^3} N^6
+\frac{C_5}{K_2^4} N^5
+\frac{C_4}{K_2^5} N^4
+\frac{C_3}{K_2^6} N^3
+\frac{C_2}{K_2^7} N^2
+\frac{C_1}{K_2^8} N^1
+\frac{C_0}{K_2^9}
=0\ ,
$$
where the coefficients $C_i$ are polynomials depending on the couplings,
on $K_2\ldots K_{8-i}$ {\it and} on $\log\left(\frac{12m^2}{K_2}\right)$.
Apart from $C_6$, that depends only on  $K_2$ quadratically,
the coefficients $C_i$ depend in general on
$K_2,\ldots K_{8-i}$, and are linear in $K_{8-i}$.
This structure thus lends itself to an iterative solution,
where killing the coefficient of $N^6$ determines $K_2$,
killing the coefficient of $N^5$ determines $K_3$ etc.
Killing all divergences determines the coefficients up to $K_7$.
%Table \ref{tab:orders} summarizes the orders to which we must expand various
%objects.
In order to calculate the coefficient $K_i$
one need to expand the equation of motion to order $n_{EOM}$ in $N$, 
which requires $\ess$ to order $n_\ess$,
which in turn requires expanding the harmonic numbers
to order $n_H$, as in the following table:
\begin{center}
\begin{tabular}{|c|c|r|r|r|r|r|r|} 
\hline
 & $K_2$ & $K_3$& $K_4$&$K_5$ &$K_6$ &$K_7$\\
\hline
$n_{EOM}$ &6&5&4&3&2&1\\
$n_\ess$ &2&1&0&-1&-2&-3\\
$n_H$&0&-1&-2&-3&-4&-5\\
\hline
\end{tabular}
\end{center}

In order to arrive at some closed expressions
we assume that $m^2$ is small compared to $\Lambda$ and $1/G$
(recall that we generally assume $\Lambda\approx 1/G$).
Then the linearized equations can be solved and in the limit $N\to\infty$ give
for the first few coefficients
\bea
K_2&=&\frac{48\sqrt{\pi\Lambda}}{\sqrt G}
+12m^2+O(m^4)\ ,
\label{firstiteration}
\\
K_3&=&-\frac{192\sqrt{\pi\Lambda}}{\sqrt G}
-48m^2+O(m^4)\ ,
\\
K_4&=&\frac{600\sqrt{\pi\Lambda}}{\sqrt G}-\frac{288\pi}{G}
+12m^2\left[
\frac{32}{3} - \frac{6 \sqrt\pi}{\sqrt{G\Lambda}} +12\xi
+(1-6\xi)
\log\left(\frac{16\pi\Lambda}{Gm^4}\right)
\right]
\ ,
\\
K_5&=&-\frac{1728\sqrt{\pi\Lambda}}{\sqrt G}
+\frac{2304\pi}{G}
+96m^2\left[
-\frac83+ \frac{6 \sqrt\pi}{\sqrt{G\Lambda}} -12\xi
-(1-6\xi)
\log\left(\frac{16\pi\Lambda}{Gm^4}\right)
\right]
\ ,
\\
K_6&=&\frac{24554\sqrt{\pi\Lambda}}{5\sqrt G}
+\frac{864\pi^{3/2}}{G^{3/2}\sqrt\Lambda}
-\frac{11808\pi}{G}
\nonumber\\
&&
+12m^2\left[
\frac{388}{15}
- \frac{247\sqrt\pi}{\sqrt{G\Lambda}} 
+488\xi
+(1-6\xi)
\left(41-\frac{6\sqrt\pi}{\sqrt{G\Lambda}}\right)
\log\left(\frac{16\pi\Lambda}{Gm^4}\right)
\right]
\ ,
\\
K_7&=&-\frac{216\sqrt{\pi\Lambda}}{5\sqrt G}(333-320\xi+960\xi^2)
-\frac{10368\pi^{3/2}}{G^{3/2}\sqrt\Lambda}
+\frac{49536\pi}{G}
\nonumber\\
&&
+48m^2\left[
\frac{116}{15}+ \frac{261\sqrt\pi}{\sqrt{G\Lambda}} -504\xi
-(1-6\xi)
\left(43-\frac{18\sqrt\pi}{\sqrt{G\Lambda}}\right)
\log\left(\frac{16\pi\Lambda}{Gm^4}\right)
\right]
\ .
\eea
The iterative procedure could of course continue determining the
coefficients of negative powers of $N$,
but that would require expanding the function $\ess$ even further.

We see that the relation $K_3=-4K_2$, that holds in the massless case,
continues to hold at linear order in a small mass.
Further, we observe that in the limit $m\to0$ the coefficients $K_{2,3,4,5}$
reduce to the ones we had already encountered in (\ref{silvia}).
However, some of the coefficients in $K_6$ and $K_7$ have a different limit.
This indicates that the limit $m\to0$ and the limit $N\to\infty$ do not commute,
at this very subleading order.

Figure \ref{fig:massive} compares the linearized iterative solution 
to a numerical solution of the full equation (\ref{cadmo}).
We see that the linear approximation is reasonably
good also for masses that are not astronomically smaller than the Planck mass.

In conclusion, we see that also a massive field will give rise to
a universe whose curvature decreases with the UV cutoff.

%%%%%%%%%%%%%%
\section{The interacting case}
%%%%%%%%%%%%%%

Let us now consider the interacting case, with kinetic operator (\ref{oper}).
The (off-shell) EA for constant $\phi$ on the sphere is
\bea
\Gamma_N(R,\phi)&=&S_H(R)+S_m(\phi,R)+\frac12\Tr_N\log\left(\frac{-\nabla^2+m^2+\tfrac12\lambda\phi^2+\xi R}{\mu^2}\right)
\nonumber\\
&=&V_4\left[\frac{\Lambda_B}{8\pi G}
-\frac{1}{16\pi G}R
+\frac12 m^2\phi^2+\frac{1}{4!}\lambda\phi^4+\frac12\xi\phi^2 R\right]
\nonumber\\
&&\quad +\frac12\sum_{\ell=1}^N m_\ell \log\left(\frac{R}{12\mu^2}\ell(\ell+3)+\frac{m^2+\tfrac12\lambda\phi^2+\xi R}{\mu^2}\right)\ ,
\label{gammaNint}
\eea
where the second line shows the classical action and the third line is the one loop trace.
The quartic self-interaction leads to a nontrivial dependence 
of the quantum corrections on $\phi$.
As in the previous sections, let us rewrite the quantum corrections in the form
\be
\frac12 \log\left(\frac{R}{12\mu^2}\right)\eff(N)
+\frac12\ti\left(N,\frac{12m^2+6\lambda\phi^2}{R}+12\xi\right)\ .
\ee
The function $\ti(N,z)$ is complex for $z<-4$.
One can see this and other properties in the left panel of Figure \ref{fig:ess}, 
showing the real and imaginary parts of $\ti(10,z)$.
Since now $z$ depends on the field $\phi$,
it means that the effective potential can become complex for some values of the field.
We will return to this point in Section 5.3.

The equation of motion of $\phi$ is
\be
\frac{64\pi^2}{R^2}\phi(6m^2+6\xi R+\lambda\phi^2)
+\frac{6\lambda\phi}{R}
\ess\left(N,\frac{12m^2+6\lambda\phi^2}{R}+12\xi\right)=0
\ ,
\label{eomphi}
\ee
while the equation of motion for the metric is
\bea
&&\frac{24\pi}{GR^3}\left(R-4\Lambda
-8\pi G\phi^2(2m^2+\xi R+\lambda\phi^2/6)\right)
\nonumber\\
&&\qquad\qquad\qquad
+\frac{1}{2R}\eff(N)
-\frac{6m^2+3\lambda\phi^2}{R^2}
\ess\left(N,\frac{12m^2+6\lambda\phi^2}{R}+12\xi\right)=0\ .
\label{eomR}
\eea
In both cases the first term is the classical one
and the second term is the quantum contribution.
The right panel of Figure \ref{fig:ess} shows a plot of $\ess(10,z)$.
Unlike $\ti$, it is everywhere real, but it has simple poles at the same locations where $\ti$ has, and is regular elsewhere.
For large $N$, these poles form a thick forest for $-N(N+3)<z<-4$.

To discuss stability we will also need the second derivatives of $\Gamma_N$.
In general one would take functional derivatives with respect to $\phi$
and $g_{\mu\nu}$, but given that we only consider $O(5)$-invariant backgrounds,
these reduce to ordinary derivatives.
In particular $\delta g_{\mu\nu}=2r\delta r\bar g_{\mu\nu}$,
where $r$ is the radius and $\bar g_{\mu\nu}$ is the metric of the unit sphere.
Then, the information about the stability is contained in the Hessian 
$$
H=\begin{pmatrix}
\frac{\partial^2\Gamma_N}{\partial\phi^2} & \frac{\partial^2\Gamma_N}{\partial\phi\partial R} \\
\frac{\partial^2\Gamma_N}{\partial R\partial \phi} & \frac{\partial^2\Gamma_N}{\partial R^2} \\
\end{pmatrix}
$$
whose matrix elements are
\be
H_{\phi\phi}=V_4
\left[m^2+\xi R+\frac12\lambda\phi^2
+\frac{\lambda R}{64\pi^2}\ess
+\frac{3\lambda^2\phi^2}{16\pi^2}\frac{\partial\ess}{\partial z}\right]\ ,
\label{secderac11}
\ee
\be
H_{\phi R}=H_{R\phi}=V_4
\left[-\frac{\phi}{3R}(6m^2+3 R\phi+\lambda\phi^2)
-\frac{\lambda \phi}{64\pi^2}\ess
-\frac{3\lambda\phi}{32\pi^2 R}(2m^2+\lambda\phi^2)\frac{\partial\ess}{\partial z}\right]\ ,
\label{secderac12}
\ee
\be
H_{R R}=V_4
\left[\frac{\phi^2(12m^2+4\xi R+\lambda\phi^2)}{4R^2}
-\frac{R-6\Lambda}{8\pi GR^2}
-\frac{\eff}{768\pi^2}
+\frac{(2m^2+\lambda\phi^2)}{64\pi^2 R}\ess
+\frac{3(2m^2+\lambda\phi^2)^2}{64\pi^2 R^2}\frac{\partial\ess}{\partial z}\right]\ .
\label{secderac22}
\ee

Depending on the shape of the effective potential, the scalar field
will have a VEV that can be zero or nonzero.
We refer to these two situations as the symmetric and the broken phase, respectively.

%%%%%%%%%%%%%%%
\section{The symmetric phase}
%%%%%%%%%%%%%%%

We begin by observing that the equation of motion of $\phi$ (\ref{eomphi})
always has a solution at $\phi=0$.
Inserting in the equation of motion of the metric (\ref{eomR}), the latter 
reduces to (\ref{cadmo}).
Thus, the solution for a free massive field discussed in Section 3 
will also be a solution for the interacting theory in the symmetric phase.
However, in the interacting case the potential receives quantum corrections.

\subsection{Stability}

The solution $\phi=0$ is not always stable:
it may be a maximum or a minimum of the potential.
Classically this is determined by the sign of $m^2$.
In the quantum theory the stability is typically determined by the
second derivative of the effective potential.
In the present situation where the metric is also dynamical,
things are a bit more complicated:
the first, third and fifth terms in the square bracket in (\ref{secderac11})
are the second derivatives of the effective potential proper,
whereas the second and fourth, that are linear in $R$,
are second derivatives of the quantum-corrected nonmininal interactions.
For the stability of the Euclidean solution what matters are 
the second derivatives of the full action, as in (\ref{secderac11}).
In the symmetric phase it simplifies to
\be
H_{\phi\phi}=V_4\left[
m^2+\xi R
+\frac{\lambda R}{64\pi^2}\ess\left(N,\frac{12m^2}{R}+12\xi\right)\right]\ .
\label{hessymm}
\ee
We have to evaluate the Hessian at the solution.
We observe that since $V_4\sim R^{-2}\sim N^4$,
the Hessian will be quartically divergent with $N$.
This, however, merely reflects the fact that in the limit $N\to\infty$ spacetime becomes flat
and its volume infinite.
What matters more is the density of the Hessian.

The function $\ess$ diverges quadratically with $N$.
However, when we evaluate the Hessian on shell,
the $R$ in the prefactor of $\ess$ goes to zero like $N^{-2}$,
removing the quadratic divergence.
Things are bit more subtle than that because the second argument of $\ess$
also depends on $N$ on shell, via the inverse of the curvature.
Clearly, what we have to do is use everywhere the ansatz (\ref{gloria})
and expand systematically.
It will be sufficient to keep the leading order of the large $N$ expansion.
Then the second argument of $\ess$ can be written $z\sim yN^2$
with $y=\tfrac{12m^2}{K_2}$ (the $\xi$ term is sub-sub-leading and can be neglected).
We can then use the expansion (\ref{Sexp}) and we find a {\it finite} density
%\be
%\frac{H_{\phi\phi}}{V_4}=
%m^2+\frac{\lambda K_2}{384\pi^2}\left[1+\frac{12m^2}{K_2}\log\left(\frac{12m^2}{K_2}\right)\right]
%+O(N^{-1})
%\label{finitehessian}
%\ee
\be
\frac{H_{\phi\phi}}{V_4}=
m^2\left\{1
+
\frac{\lambda}{32\pi^2}\left[\frac{K_2}{12m^2}+\log\left(\frac{12m^2}{K_2+12m^2}\right)\right]
+O(N^{-1})\right\}
\label{finitehessian}
\ee
where $K_2$ is given by (\ref{firstiteration}).
Recalling that $0<m^2<K_2$,
this matrix element of the Hessian is always positive.
Thus, for $m^2>0$ we are in the symmetric phase,
also taking into account the quantum corrections.

For $\phi=0$ the off-diagonal elements $H_{R\phi}$ and $H_{\phi R}$ are zero.
There remains the $RR$-element, that simplifies to
\be
H_{R R}=V_4
\left[
-\frac{R-6\Lambda}{8\pi GR^2}
-\frac{\eff}{768\pi^2}
+\frac{m^2}{32\pi^2 R}\ess
+\frac{3m^4}{16\pi^2 R^2}\frac{\partial\ess}{\partial z}\right]
\label{secderac22b}
\ee
We have seen that for large $N$,
$\ess(N,y N^2)\sim N^2$.
The derivative of $\ess$ is
\be
\ess'(N,z)\equiv\frac{\partial\ess(N,z)}{\partial z}
=-\sum_{\ell=1}^N m_\ell
\frac{1}{(\ell(\ell+3)+z)^2}\ .
\label{essp}
\ee
Since the numerator in the sum is cubic, and the denominator is quartic in $N$,
one expects the sum to diverge logarithmically with $N$.
However, when we go on shell, the second argument of $\ess'$ diverges as $z\sim y N^2$
and evaluating the function in this limit gives the result (\ref{essprime}).
Thus we find that $\ess'$ is finite, and overall each of the three quantum terms 
in (\ref{secderac22b}) is quartically divergent.
The coefficient of this divergence can be estimated for small $m$
and is positive.
Thus the solution is stable also in the direction of gravitational perturbations.
For $m^2=0$ this can also be gleaned from Figure \ref{fig:gamma}.

\subsection{Renormalization}

The density of the $H_{\phi\phi}$ can be interpreted as a quantum-corrected
effective mass, and therefore the preceding calculation also leads to the 
important physical conclusion that the quantum correction to the mass is finite.
In order to appreciate this point, let us compare the preceding calculation
with a more standard quantum field theoretical procedure.
Normally one would treat $\phi$ and $R$ as being externally fixed,
and look for the dependence of the Hessian on the cutoff.
Thus we would expand (\ref{hessymm}) for large $N$ keeping $R$ fixed,
leading to
\be
\frac{H_{\phi\phi}}{V_4}
\sim
m^2+\xi R
+\frac{\lambda R}{64\pi^2}\bigg[
\frac16 N^2
+\frac23 N
-\frac{2(6m^2+(6\xi-1)R)}{3R}\log[N]
\bigg]+\mathrm{finite\ terms}\ .
\label{hessymmdiv}
\ee
Then, one could define the quantum-corrected mass and nonminimal coupling by
\bea
m_{eff}^2&=&\frac{1}{V_4}
\frac{\partial^2\Gamma_N}{\partial\phi^2}
\bigg|_{\phi=0,R=0}
\\
\xi_{eff}&=&\frac{1}{V_4}
\frac{\partial^3\Gamma_N}{\partial R\partial\phi^2}
\bigg|_{\phi=0,R=0}
\eea
and one sees from (\ref{hessymmdiv}) that these quantities are divergent, namely
\bea
m^2_{eff}&=&m^2-\frac{\lambda}{16\pi^2}m^2\log N+O(N^0)
\label{meff}
\\
\xi_{eff}&=&\xi+\frac{\lambda}{64\pi^2}
\left[
\frac16 N^2+\frac23 N-4\left(\xi-\frac16\right)\log[N]
\right]+O(N^0)\ .
\label{xieff}
\eea
We recognize here the standard logarithmic divergence of the mass
that is proportional to the mass itself,
and the fact that the log divergence in $\xi$ vanishes in the conformal case.
When one compares this to the standard results obtained with a dimensionful
momentum cutoff (see Appendix A)
there is only one unusual feature, namely the power divergences,
that normally affect the mass, appear here in $\xi$ instead.

This slight oddity, however, is due to the fact that our cutoff is dimensionless,
and disappears when we use (\ref{ddcut}) to convert the dimensionless cutoff $N$
to a dimensionful cutoff $C$.
If we do this, (\ref{hessymmdiv}) becomes
\be
\frac{H_{\phi\phi}}{V_4}
\sim
m^2+\xi R
+\frac{\lambda}{32\pi^2}\bigg\{
\cut^2
+\frac{2}{\sqrt3} \cut\sqrt R
-m^2\log[C^2/R]
-\left(\xi-\frac16\right)R\log[C^2/R]
\bigg\}+\mathrm{finite\ terms}\ ,
\label{hessymmdiv2}
\ee
which, aside from the linearly divergent term and a finite additive constant,
is identical to (\ref{divcut}).
Whether we look at it in this form, or with the dimensionless cutoff as in (\ref{hessymmdiv}),
this result shows the presence of divergences,
that one would normally absorb in the definition
of a renormalized mass and nonminimal coupling.
However, this is unnecessary when we put the metric on shell.
Since then $R\sim K_2/N^2$, the first term in the bracket in (\ref{hessymmdiv}) becomes finite, 
the second goes to zero,
the coefficient of $12m^2\log N$ becomes finite, equal to $\tfrac{\lambda K_2}{384\pi^2}$, 
and the coefficient of $(6\xi-1)\log N$ goes to zero.
We seem to remain just with the logarithmic divergence that renormalizes the mass.

However, let us have a more careful look at the finite terms in (\ref{hessymmdiv}).
They are given by
\be
\mathrm{finite\ terms}=
\frac{(6m^2+(6\xi-1)R)}{3R}
\left[
\psi_0\left(\frac{5+\sqrt{9-48\frac{m^2+\xi R}{R}}}{2}\right)
+\psi_0\left(\frac{5-\sqrt{9-48\frac{m^2+\xi R}{R}}}{2}\right)
\right]
\ee
where $\psi_0$ are polygamma functions.
When we set $R=K_2/N^2$ and expand for large $N$,
using that $\psi_0(x)\sim\log(x)$ for $x\to\infty$,
we see that the apparently finite terms actually
have a logarithmic divergence that exactly cancels the one in (\ref{hessymmdiv}),
leaving behind just a finite term
$$
\frac{\lambda}{32\pi^2}m^2\log\left(\frac{12m^2}{K_2}\right)\ .
$$
When $R$ is on shell, it becomes actually impossible to separate
the mass from the nonminimal interaction,
so defining an effective mass as the sum of these two terms, on shell,
the preceding calculation leads to
\be
\tilde m_{eff}^2\equiv\frac{1}{V_4}
\frac{\partial^2\Gamma_N}{\partial\phi^2}
\bigg|_{\phi=0,R=R_*}
=m^2+\frac{\lambda}{384\pi^2}\left[K_2+12m^2\log\left(\frac{12m^2}{K_2}\right)\right]
+O(N^{-1})
\label{finitemass0}
\ee
where $R_*$ is the solution.
However, the contribution of $\xi R$ on shell is negligible,
so this can be rightfully be seen as the quantum-corrected mass.

This is not equal to (\ref{finitehessian}) for the following reason.
In addition to the terms of order $N^0$ that we have already considered,
the expansion (\ref{hessymmdiv}) contains infinitely many terms with inverse
powers of $N$, that also contain $R$.
The terms with odd powers of $1/N$ are of the form $1/(N^{2k+1}R^k)$
and therefore go to zero for $N\to\infty$, whereas 
terms with even powers of $1/N$ are of the form $1/(N^{2k}R^k)$
and leave a finite contribution.
Resumming all these contributions makes up the difference
between the partial result (\ref{finitemass0}) and the correct full result (\ref{finitehessian}).
%(We discuss this resummation for the function $\ess$ in Appendix XXX.)
The only advantage of this alternative route is to make connection with the standard
approach and to see in detail how the cancellation of divergences works.
\smallskip

The renormalization of the  quartic self-interaction works in a similar way.
The quantum-corrected  $\lambda$ is given by
\be
\lambda_{eff}\equiv\frac{1}{V_4}\frac{\partial^4\Gamma_N}{\partial\phi^4}
\bigg|_{\phi=0}
=\lambda+\frac{9\lambda^2}{16\pi^2}\ess'\left(N,\frac{12(m^2+\xi R)}{R}\right)\ .
\label{lambdadiv}
\ee
Let us follow the standard procedure and expand $\ess'$ for large $N$ and fixed $R$.
As we saw in (\ref{essp}), $\ess'(N,z)$ grows as $\frac13\log[N]+O(N^0)$,
so {\it off shell} we find the familiar logarithmic divergence
\be
\lambda_{eff}=\lambda+\frac{3\lambda^2}{16\pi^2}\log[N]+O(N^0)\ .
\label{lambdalog}
\ee
This is in perfect agreement with the standard
calculation with dimensionful cutoff (\ref{quartic}).
However, let us proceed also in this case by a more careful evaluation
keeping also the terms of order $O(N^0)$. They are given by
\bea
%\lambda_{eff}\!\!&=&\!\!
%\lambda
%-\frac{3\lambda^2}{16\pi^2}\log N
%\nonumber\\
%
&&
\frac{3\lambda^2}{32\pi^2}
\left[
\psi_0\left(\frac{5+\sqrt{9-48\frac{m^2+\xi R}{R}}}{2}\right)
+\psi_0\left(\frac{5-\sqrt{9-48\frac{m^2+\xi R}{R}}}{2}\right)
\right]
%\nonumber
\\
&&
+\frac{\sqrt3\lambda^2(6m^2+(6\xi-1)R)}{16\pi^2\sqrt{-16m^2 R+(3-16\xi)R^2}}
\left[
\psi_1\left(\frac{5+\sqrt{9-48\frac{m^2+\xi R}{R}}}{2}\right)
+\psi_1\left(\frac{5-\sqrt{9-48\frac{m^2+\xi R}{R}}}{2}\right)
\right]\ .
\nonumber
\eea
When we put the gravitational field on shell by using (\ref{gloria}),
the arguments of the polygamma functions becomes of order $N$
and we have, for large $N$, $\psi_0\sim \log N$, $\psi_1\sim 1/N$.
However, the coefficient of $\psi_1$ is of order $1/\sqrt R$,
so those terms give finite contributions, whereas the $\psi_0$ terms
cancel the explicit $\log N$ in (\ref{lambdalog}).
What remains is a finite renormalization of $\lambda$:
\be
\lambda_{eff}=\lambda+\frac{3\lambda^2}{32\pi^2}
\left[1+\log\left(\frac{12m^2}{K_2}\right)\right]\ ,
\label{finitel0}
\ee
where $K_2$ is given by (\ref{firstiteration}).
As in the calculation of $\tilde m_{eff}^2$,
this shows that the divergences cancel, but it does not give the correct finite part.
This is because the terms with inverse powers of $N$ also contain
inverse powers of $R$ that, on shell, make them finite.
The correct result can be obtained most straightforwardly
by using directly the on-shell expansion (\ref{essprime})
in (\ref{lambdadiv}):
\be
\lambda_{eff}=\lambda+\frac{3\lambda^2}{32\pi^2}
\left[\frac{K_2}{12m^2+K_2}+\log\left(\frac{12m^2+K_2}{K_2}\right)\right]\ .
\ee

%%%%%%%%%%%%%%%%%%%%%%
%%%%%%%%%%%%%%%%%%%%%%
\section{Broken phase}
%%%%%%%%%%%%%%%%%%%%%%
%%%%%%%%%%%%%%%%%%%%%%

It is well known that strictly speaking phase transitions can occur only
in the limit of infinite volume and therefore are not possible on spherical spacetimes.
However, we have seen that when the metric is required to solve the semiclassical
Einstein equations, its volume goes to infinity in the limit $N\to\infty$.
There is thus a chance that phase transitions may occur in this limit.
In this section we give some preliminary results on this issue.
We shall see that starting with a symmetry-breaking classical potential,
also the perturbative effective potential can have nontrivial vacua.
However, this is not a conclusive argument due to various hurdles,
some of which already arise in flat spacetime \cite{Weinberg:1987vp}, 
while others are peculiar to the spherical geometry \cite{Benedetti:2014gja}.
In particular, it is well-known that the effective potential must be convex,
a condition that is violated by the perturbative result,
and that it is complex in the region between the two inflection points.
The perturbative effective potential gives the energy of a homogeneous
field configuration, but the minimum of the energy for fields between the two
minima is a nonhomogeneous configuration and is energetically degenerate
with the minima.
In our calculations, depending on the values of the parameters,
the effective potential can become complex even outside the minima.
We are going to avoid such situations.

On the sphere we encounter some additional complications.
To see this, consider the arguments of the logs in the formula for the EA:
\be
\frac{R}{12}\ell(\ell+3)+m^2+\frac12\lambda\phi^2\ ,
\label{sings}
\ee
where we have put $\xi=0$ for simplicity.
We consider a fixed large value of $N$.
When $m^2<0$ there is the danger of the argument of the log becoming negative.
This danger is greates for the lowest eigenvalue, so let us focus first on the mode $\ell=1$.
The argument of the first logarithm,
$\tfrac{R}{3}+m^2+\frac12\lambda\phi^2$
is positive for any value of $\phi$ as long as $-R/3<m^2<0$ 
but if $m^2<-R/3$ it becomes zero for
$\phi^2=\frac{2}{\lambda}\left(|m^2|-\frac{R}{3}\right)\equiv\phi_1^2$
and negative for $\phi^2<\phi_1^2$.
So the effective potential has a singularity at $\pm\phi_1$
and is complex for $\phi^2<\phi_1^2$.
The second logarithm gives another singularity at 
$\phi_2^2=\tfrac{2}{\lambda}\left(|m^2|-\frac{5R}{6}\right)<\phi_1^2$
and an additional imaginary contribution for $\phi^2<\phi_2^2$.
Each mode gives rise to a pole and an imaginary contribution
until eventually $\tfrac{R}{12}\ell(\ell+3)$ becomes sufficiently large
to offset the negative $m^2$.

If $N$ is greater than this value of $\ell$, then increasing $N$ at fixed $R$
will not add further poles.
However, we want to keep $R$ on shell.
If $R\approx N^{-2}$ as in the symmetric phase,
and we will see that this is the case,
then increasing $N$ can increase the number of logs with negative argument.
In fact, for $N\to\infty$, we have $R\to 0$, so the first logarithm
has a singularity at
\be
\phi^2=-\frac{2m^2}{\lambda}\ .
\label{appsing}
\ee
Keeping the parameters fixed and taking the limit $N\to\infty$, 
the poles form a dense forest for $\phi^2<2|m^2|/\lambda$.
The problem is that the stationary points of the effective potential
may occur {\it inside} this region, and in this case the physical meaning
of the solution is obfuscated.
We will say that the theory is in the broken phase if the effective potential
has nontrivial minima in the region where the effective potential is real.

%%%%%%%%%%%%%%%%
\subsection{Approximate solution}

Next we look for solutions of the equations  (\ref{eomphi}), (\ref{eomR})
with $\phi\not=0$.
In this case we can eliminate $\ess$ from these two equations
and obtain a simpler equation:
\be
\frac{24\pi}{G R^3}
\left(-R+4\Lambda
-\frac{16\pi Gm^2}{3\lambda}(3m^2+3\xi R-\lambda\phi^2)
\right)
=\frac{1}{2R}\eff(N)
\label{eommix}
\ee
This is very similar to (\ref{eommassless}),
except that the l.h.s. now depends on the scalar.
It is convenient to replace the system (\ref{eomphi}-\ref{eomR}),
where the function $\ess$ appears twice,
with the system (\ref{eomphi}-\ref{eommix}),
where the function $\ess$ appears only once.

The equations can be solved numerically,
but they can also be solved analytically if one makes some additional approximation.
Motivated by some preliminary numerical investigations,
we begin with the ansatz (\ref{gloria}),
supplemented by the assumption that the VEV of the scalar
is independent of $N$ in the large $N$ limit:
\be
R=\frac{K_2}{N^2}+\frac{K_3}{N^3}+\ldots\ ,\qquad
\phi= F_0+\frac{F_2}{N^2}+\ldots\ .
\label{leading}
\ee
Due to the complication of the broken phase, in the following we will
limit ourselves to the leading term of each expansion.
This means that the solution for $z$ must have the form (\ref{zy}),
where $y$ is a constant.
The nonminimal coupling $\xi$ can be ignored in this leading order.
We solve (\ref{eommix}) for $R$, insert in (\ref{eomphi}) together with the ansatz,
expand for large $N$ and
extract the coefficient of the highest power of $N$, namely $N^2$.
Demanding that this coefficient be zero leads to the following equation for $y$:
\be
-\frac{8 (56\pi^3Gm^4  + 6 \pi^2 \lambda \Lambda)}{9 \lambda(3 \lambda \Lambda - 20\pi G m^4 )}y
-\frac{8\pi m^2\sqrt{\pi G (4 G m^4 \pi (64 \pi^2 y^2 - 15 \lambda) + 
		9\lambda^2 \Lambda)}}{9\lambda(3 \lambda \Lambda - 20\pi G m^4)}
=\frac16\left[1-y\log\left(1+\frac{1}{y}\right)\right]
\ee
This equation cannot be solved in closed form,
but can be solved graphically by intersecting the
line on the l.h.s. with the graph of the function on the r.h.s.
When the intersection occurs for $y\ll1$, the logarithmic term can be neglected
and the r.h.s. becomes equal to the constant 1/6.
In this regime the solution can be approximated by
\bea
y&=&
%\frac{-A_1^2-96\pi A_3 A_4
%\pm A_1\sqrt{A_1^2(1-12A_3)-576\pi  A_3^2A_4}}{12(A_1^2A_3+48\pi A_3^2 A_4)}
%\nonumber\\
%&=&
\frac{1}{32\pi^2}
\frac{3\lambda^2\Lambda-28\pi G\lambda m^4
\pm8\pi m^2\sqrt\lambda\sqrt{4\pi G\lambda\Lambda-m^4G^2(48\pi^2-\lambda)}}{12\pi G m^4-\lambda\Lambda}
\label{approxsol}
\eea

From $K_2=R N^2$, using the solution for $R$ and (\ref{eff}) we obtain to leading order
\be
K_2=6\left[\frac{128\pi^2 m^4 G}{\lambda} y 
\pm\sqrt{\left(\frac{128\pi^2 m^4 G}{\lambda}\right)^2  y^2-16\pi \left(\frac{80 \pi m^4}{3\lambda}-\frac{4\Lambda}{G}\right)}\right]\ .
\label{alicek}
\ee
Inserting in this formula the approximate solution (\ref{approxsol}) we obtain 
a complicated expression for $K_2$.
Finally from (\ref{zy}) and solving the formula for $z$ (with $\xi\to0$ and $R\to K_2/N^2$)
for $\phi$, we obtain the constant value of $\phi$:
\be
F_0^2=\frac{y K_2-12m^2}{6\lambda}\ ,
\label{F0}
\ee
that should be compared to the minimum of the classical potential at $\phi^2=-6m^2/\lambda$.

When one uses (\ref{approxsol}) to write $K_2$ and $F_0$
one obtains very long expressions that we do not report here,
but that can be easily plotted.
In any case, assuming that the solution is within the domain of the approximations,
we have thus shown that in the broken phase it is self-consistent to assume
that in the limit $N\to\infty$, $R\sim N^{-2}$ and $\phi\sim$constant.

\begin{figure}[]
\centering
\includegraphics[width=.5\textwidth]{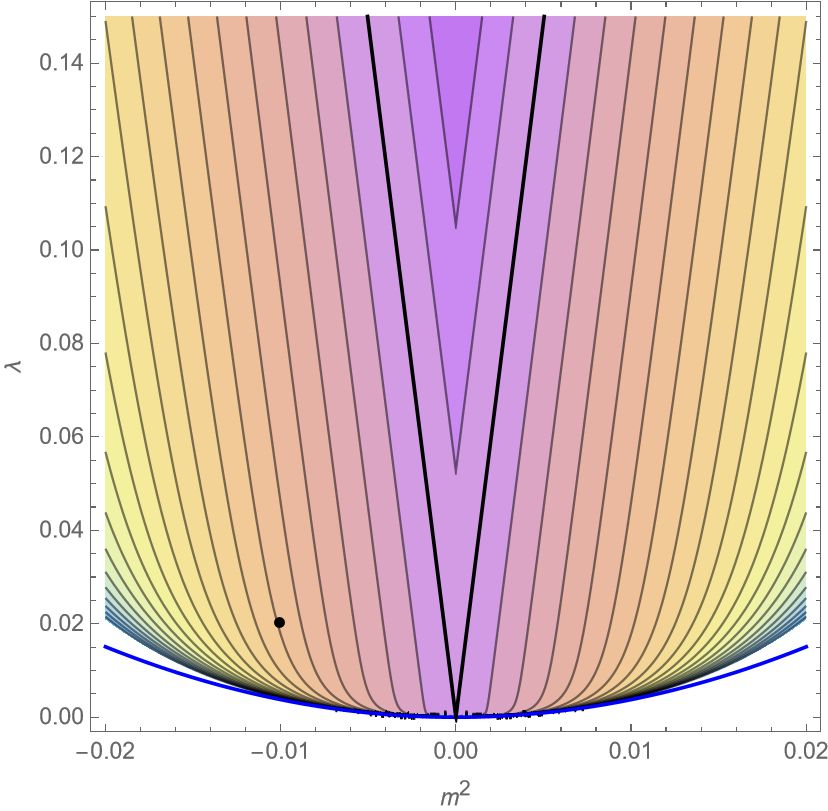}
\caption{Contour plot of the solution (\ref{approxsol}) for $\Lambda=1$,
as function of $m^2$ and $\lambda$.
The solution has a singularity near the boundary
of the colored region (blue curve in the bottom) and becomes complex outside/below it.
The black V is the $y=0$ level curve.
The function is positive outside the V and negative inside.
The distance between the level curves is $0.0005$.
The dot is the reference point.
}
\label{fig:ypsilon}
\end{figure}

To get a feeling of the parameter space where the approximation of neglecting
the log term may be valid, we show in Figure \ref{fig:ypsilon} the function $y(m^2,\lambda)$.
It is zero on the V-shaped level curve given by
\be
\lambda=\frac{2\pi m^2G}{3\Lambda}\left(5m^2\pm\sqrt{25m^4+\frac{64\pi\Lambda}{G} }\right)
\ee
so we expect the approximation to be good in a neighborhood of that curve.
On the other hand, if we let $\lambda$ tend to zero for fixed $m^2$
or $m^2$ grow for fixed $\lambda$, the solution hits a singularity.

In order to establish the stability of the solution,
in the leading large $N$ approximation Equation (\ref{secderac11}) reduces to
\be
\frac{H_{\phi\phi}}{V_4}\approx\frac{1}{36}\left(\frac{18\lambda}{32\pi^2}\right)^2
\left(-\frac{128 \pi^2 m^4 G}{\lambda}+\frac{64\pi^2}{18\lambda} K_2 y\right)\left(\frac{32\pi^2}{18\lambda}+\frac{\partial\ess}{\partial z}\right)\ .
\label{secderacz11onshell2}
\ee
In the approximate solution one can insert the formulas (\ref{approxsol}) and (\ref{alicek})
for $y$ and $K_2$ and obtain a very complicated but completely explicit formula.

%%%%%%%%%%%%%%%%%%%%%%%%
\subsection{Reference point}

We can compare the approximate solution to a numerical solution
of the full equations, for some specific values of the couplings.
For example, let us choose
$$
\Lambda G=1\ ,\quad
m^2 G=-0.01\ ,\quad
\lambda=0.02\ ,\quad
\xi=0\ .\quad
$$
%that corresponds to
%$$
%A_1\approx -210\ ,\quad
%A_2=0\ ,\quad
%A_3\approx 877\ ,\quad
%A_4\approx -3.58\ ,\quad
%A_5=1\ .\quad
%$$
Inserting these values in (\ref{approxsol}) we obtain $y=0.00293$,
so this point should lie in the domain of the approximation.

\begin{figure}[]
\centering
\includegraphics[width=.48\textwidth]{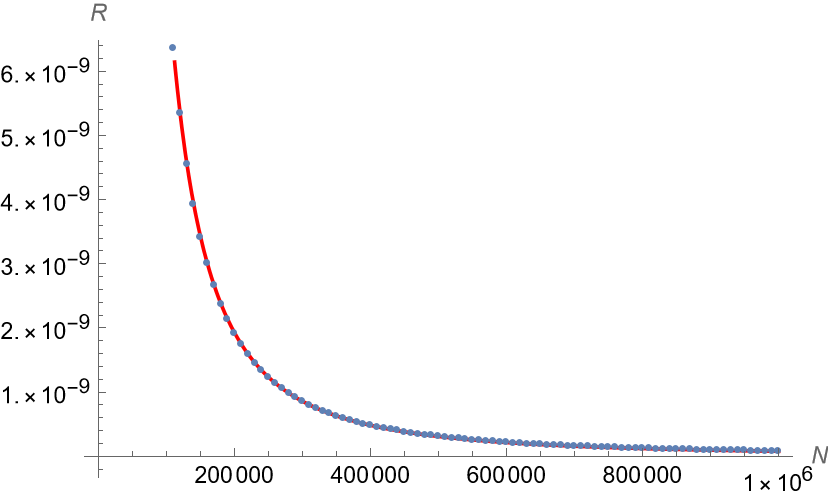}
\ \ \
\includegraphics[width=.48\textwidth]{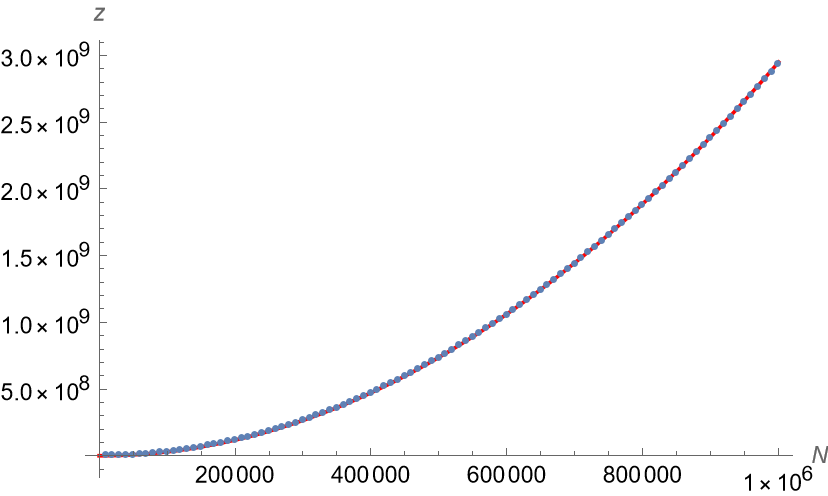}
\caption{Numerical solution of the equations of motion for $R$ (left) and
the corresponding value of $z$ (right),
with $10000<N<1000000$.
The plot on the left is superimposed on the curve $R=K_2/N^2$, giving an estimate
of the parameter $K_2\approx 77$.
The plot on the right is superimposed on the curve $z=y N^2$, giving an estimate
of the parameter $y\approx 0.00295$.
}
\label{fig:numsol}
\end{figure}

Figure \ref{fig:numsol} shows the numerical solution of equations (\ref{eomphi},\ref{eomR})
for increasing $N$.
It fits perfectly the leading behavior given in (\ref{leading}),
for the parameter values $K_2\approx 77$ and $F_0=1.697$
(we do not plot $\phi$ as a function of $N$ because it is constant 
within 10 decimal places, over the whole range).
For these values of the couplings, one can also compare the approximate
formula for the effective mass (\ref{secderacz11onshell2})
to the full formula (\ref{secderac11}), and again we observe excellent agreement.
One can see numerically that the mass converges very rapidly to a finite value as a function of $N$.

In the following table we compare the numerical results 
for the reference point with $N=10^6$, 
to the approximate solution of the preceding section and we find very good agreement.

\begin{center}
\begin{tabular}{|c|l|l|l|l|} 
\hline
 & $y$ & $K_2$ & $F_0$ & $m_*^2$\\
\hline
full numerical & 0.00295  & 77 & 1.6972 & 0.01916 \\
approximate analytic & 0.00293 & 76.9 & 1.6965 & 0.01919\\
\hline
\end{tabular}
\end{center}

%%%%%%%%%%%%%%
\subsection{Numerical solutions}

Having found an solution of the equation in some range of parameter space,
we will now try to get some sense for how far the solution extends.
For that we shall probe numerically some directions in parameter space.
A systematic exploration is beyond the aim of this paper and will be left for the future.

In view of possible realistic applications,
we are mainly interested in the limit $Gm^2\to 0$.
We begin by taking this limit, keeping all other parameters fixed.
Starting from the reference point of Section 5.2, that has $m^2=-1/100$,
we decrease the modulus of $m^2$ and observe that the value of $R$ decreases
by a fractionally very small amount, while $\phi$ tends towards zero.
This is the expected behavior, since the classical potential has a phase transition
at $m^2=0$ and one expects the VEV of $\phi$ to be zero for $m^2>0$.
However, the solution ceases to exist at $m^2\approx -0.0007$.
The reason for this is that the solution enters the region where
the potential is complex.

\begin{figure}[]
\centering
\includegraphics[width=.8\textwidth]{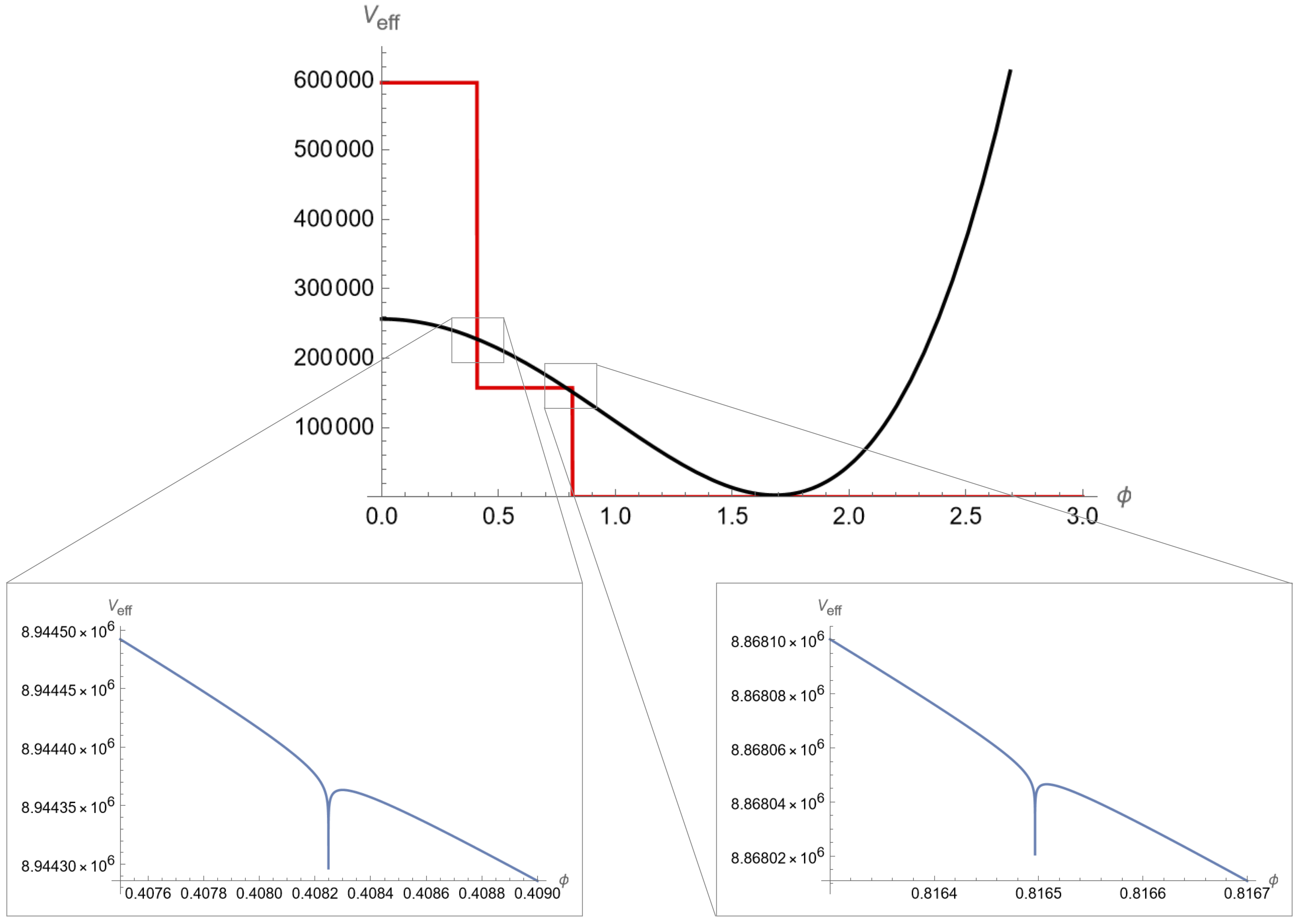}\\
%\includegraphics[width=.4\textwidth]{pole2.png}
%\ \ \ 
%\includegraphics[width=.45\textwidth]{pole1.png}
\caption{The effective potential at the reference point and $N=100$.
Black curve: real part of the potential, shifted by $-8.717\times 10^6$;
red curve: imaginary part of the potential, multiplied by $20000$.
The real part of the potential looks smooth, but it has two poles,
as shown by the enlargements below.
}
\label{fig:V100}
\end{figure}

\begin{figure}[]
\centering
\includegraphics[width=.47\textwidth]{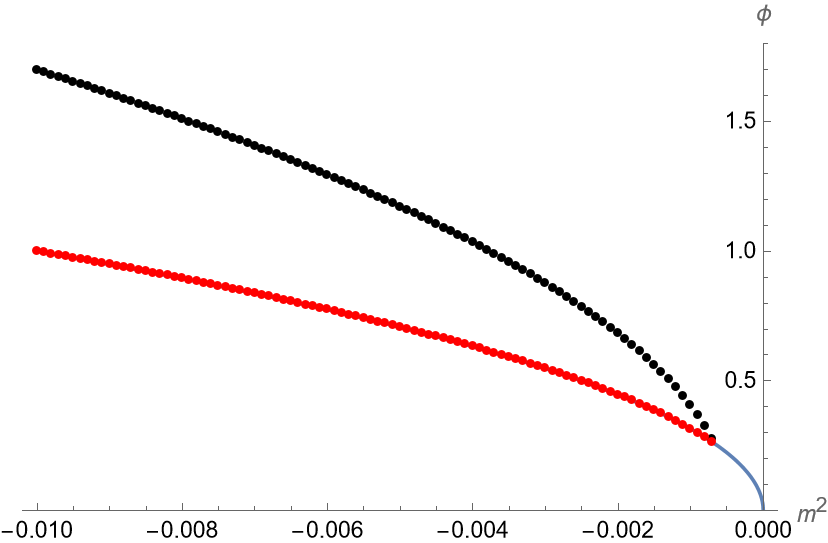}
\ \
\includegraphics[width=.47\textwidth]{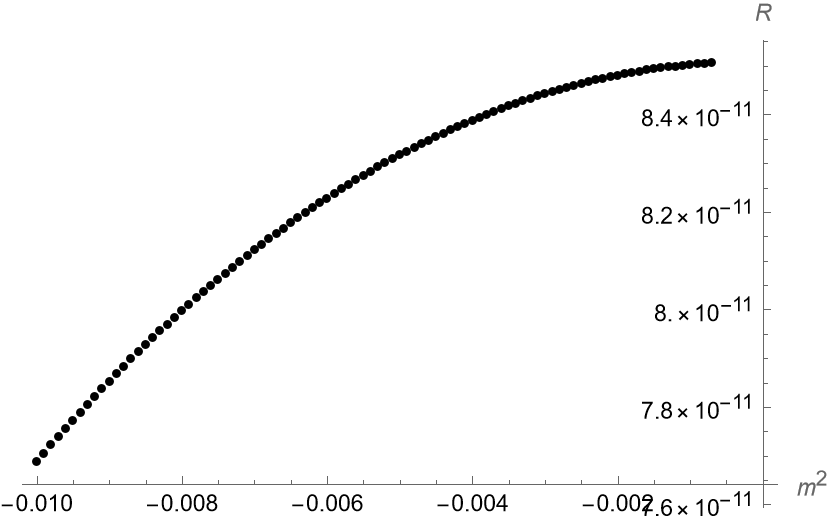}
\caption{Numerical solution starting from the reference point on the left and decreasing
the absolute value of $m^2$. Left, black: position of the minimum of the effective potential 
as function of $m^2$.
Left, red: position of the pole of the first logarithm, superimposed on the curve (\ref{appsing}).
Right: solution for $R$ as function of $m^2$.
All  with $G=\Lambda=1$, $\xi=0$, $\lambda=1/50$, $N=10^6$.
}
\label{fig:solpole}
\end{figure}

At some large and fixed $N$, we see from (\ref{sings}) that decreasing $m^2$
has the effect of decreasing the value of $\phi$ where the first pole occurs.
For sufficiently large $N$ the first term in (\ref{sings}), with $\ell=1$, is negligible,
and the first pole occurs at (\ref{appsing}).
However, decreasing $m^2$ also decreases the value of $\phi$ where the minimum
of the potential occurs, and the minimum moves faster than the pole,
so that eventually the poles fall outside the minima,
and the solution falls in the region where
the effective potential is complex. See Fig. \ref{fig:solpole}.
Thus we find that the theory has a symmetric phase for $m^2>0$
and a broken phase for $m^2$ more negative than some critical value.
For $m^2<0$ but greater than this critical value,
the potential at the solution is complex.
Thus there is no continuous transition between the two phases.

A different way of reaching small masses is to decrease $|m^2|$ and $\lambda$ at the same rate.
In the classical potential, such a limit keeps the VEV of $\phi$ constant.
If we start again from the reference point and send $m^2$ to zero in this way,
the numerical solution does not seem to encounter any obstacle.
In fact, the location of the first pole is always fixed at the value (\ref{appsing}),
which is numerically $\sim 1$,
but the solution is also almost constant and has a finite limit $\phi\approx 1.6927$.
That this should be the case can also be seen analytically.
In fact, looking at  (\ref{approxsol},\ref{alicek},\ref{F0})
we see that for $m^2\sim\lambda\sim\epsilon\to 0$ we have
$y\sim\epsilon$, $K_2\sim\epsilon^0$ and $F_0\sim\epsilon^0$.

\vskip1cm

%%%%%%%%%%%%%%
\subsection{Renormalization}

Having established that for $N\to\infty$ the solution in the broken phase,
at least in some region of parameter spaces, has the
leading behavior  $R\sim K_2/N^2$ and $\phi\sim F_0$,
we can look at the renormalization of the mass and quartic coupling,
defined as derivatives of the effective potential at the nontrivial minimum.
Since $\phi\ne 0$, there are now new terms compared to the symmetric phase.
For the mass we have
\bea
\tilde m_{eff}^2\equiv\frac{1}{V_4}
\frac{\partial^2\Gamma_N}{\partial\phi^2}
\bigg|_{\phi=\phi_*,R=R_*}
&=&m^2+\xi R
+\frac{\lambda R}{64\pi^2}\ess\left(N,\frac{12m^2+6\lambda\phi^2}{R}+12\xi\right)
\nonumber\\
&&
\qquad\qquad+\frac{3\lambda^2\phi^2}{16\pi^2}\ess'\left(N,\frac{12m^2+6\lambda\phi^2}{R}+12\xi\right)\ .
\eea
Since $\langle\phi\rangle\sim F_0$ is independent of $N$ for $N\to\infty$,
the arguments of $\ess$ and its derivative are the same as in the symmetric phase,
except for the replacement of $12m^2$ by $12m^2+6\lambda F_0^2$.
We have already seen in (\ref{finitehessian}) that $R\, \ess(N,y N^2)$ is finite
and in (\ref{essp}) that and $\ess'(N,y N^2)$ is finite.
Then, by the same arguments used in the symmetric phase,
the quantum correction to the mass is finite on shell.

The quantum-corrected  $\lambda$ is given by
\bea
\lambda_{eff}&\equiv&\frac{1}{V_4}\frac{\partial^4\Gamma_N}{\partial\phi^4}
\bigg|_{\phi=\phi_*}
=\lambda
+\frac{9\lambda^2}{16\pi^2}\ess'\left(N,\frac{12m^2+6\lambda\phi^2}{R}+12\xi\right)
%\nonumber
\\
&&
+\frac{27\lambda^3\phi^2}{2\pi^2 R}\ess''\left(N,\frac{12m^2+6\lambda \phi^2}{R}+12\xi\right)
%\nonumber\\
%&&
+\frac{27\lambda^4\phi^4}{\pi^2 R^2}\ess'''\left(N,\frac{12m^2+6\lambda\phi^2}{R}+12\xi\right)\ .
\nonumber
\eea
The first term is finite, as we have just seen, and the remaining two terms contain
\bea
\ess''(N,z)&=&2\sum_{\ell=1}^N m_\ell\frac{1}{(\ell(\ell+3)+z)^3}
\\
\ess'''(N,z)&=&-6\sum_{\ell=1}^N m_\ell\frac{1}{(\ell(\ell+3)+z)^4}
\eea
that are convergent even before going on shell.
Thus, the renormalization of $\lambda$ is finite on shell also in the broken phase.

\vskip1cm

%%%%%%%%%%
\section{Discussion}
%%%%%%%%%%

Let us summarize our main points.
For a {\it self-interacting} scalar field $\phi$ in a Euclidean De Sitter space,
the spectrum is discrete and it is natural to cut off the sum over modes at some
value $N$ of the principal quantum number $\ell$.
The total number of modes that is kept in this way is $O(N^4)$.
Then:
\\
(1) 
In the symmetric phase,
solving the semiclassical Einstein equations and the equation for $\phi$, 
that come from varying the quantum EA,
the De Sitter radius is found to grow linearly with $N$ for large $N$,
while the VEV of the scalar is zero.
\\
(2) In the symmetric phase, the mass and the quartic coupling of the scalar,
defined as the second and fourth derivative of the EA
at the solution of the above mentioned equations of motion,
receive only a finite renormalization when $N\to\infty$.
\\
(3) We have found evidence for the existence of a broken phase when $m^2$
is sufficiently negative. This phase seems to be separated from the
symmetric phase by a region where the potential at the solution is complex.
These results will have to be confirmed by further analyses
and the meaning of the complex region will have to be investigated.
In the broken phase, as in the symmetric one, 
the De Sitter radius grows linearly with $N$,
the VEV of the scalar is independent of $N$ and the renormalizations of the
mass and coupling are finite.
\smallskip

There are two main features in the calculations that lead to these results.
The first is the use of a dimensionless cutoff, the second is
the use  of the equation of motion for $R$.
In the case of a sphere $S^4$, where the spectrum of the Laplacian is discrete,
the use of a dimensionless cutoff is the most natural option and is closely related to ideas
in noncommutative geometry \cite{Madore:1991bw,Fiore:2017ude}.
\footnote{It is amusing to note that, as in noncommutative geometry, 
it may be possible to see in our calculations a form of UV/IR mixing: 
the limit $N\to\infty$ is certainly the UV limit of the theory, 
but it gives rise, via the gravitational equations of motion,
to an infinite volume (an IR divergence).}
One may think that there cannot be an essential difference between a dimensionless cutoff
and a standard one with dimension of momentum, since they can be related as in (\ref{ddcut}).
This would indeed be the case if we treated $R$ as a fixed parameter
as in the comparison between (\ref{hessymmdiv}) and (\ref{hessymmdiv2}).
There we saw that the usual quadratic divergence of the mass gets reinterpreted,
with the dimensionless cutoff, as a divergence of the nonminimal coupling
and what remains is just the logarithmic divergence, proportional to the mass itself.
This is to be expected of a dimensionless regulator,
and is similar to what happens e.g. in dimensional regularization.
However, as long as one treats $R$ as an arbitrary externally given parameter,
there are divergences independently of the dimension of the cutoff.

This is why the second feature is necessary:
the equation for the metric has to be solved {\it before} sending the cutoff to infinity.
In the calculation of $R$, this is what produces a decrease of $R$ with $N$.
In the calculation of the EA, it is by putting the metric on shell
that the quantum corrections to the mass and quartic coupling become finite.
The cancellation of divergences could be seen in part as a consequence of dimensional
analysis, due to our choice for the dimension of the cutoff.
Indeed, the quadratic ($N^2$) divergence in (\ref{hessymmdiv}) is seen to
be canceled {\it multiplicatively} by the on shell $N$-dependence of the prefactor $R$.
However, the logarithmic divergences, both in (\ref{hessymmdiv}) and in (\ref{lambdadiv})
are canceled {\it additively} when one puts the metric on shell, 
indicating that this is a more subtle effect.

We emphasize that both features are necessary to arrive at the results.
We have already seen that if we treat $\phi$ and especially $R$
as externally fixed backgrounds, then the curvature is an increasing
function of $N$ and the sums over $N$ defining the effective mass 
and quartic coupling are divergent.
Thus, the use of a dimensionless cutoff is not enough
and the necessity of going on shell is clear.
This is the feature that in \cite{Becker:2020mjl,Pagani:2019vfm,Ferrero:2022hor} 
was called {\it background independence}:
The necessity of using a dimensionless cutoff may be less obvious at this point.
For this reason it is instructive to consider an alternative point of view,
that we have described in Appendix A.
We start from the usual way of calculating the divergences in curved spacetime,
based on the small-time (asymptotic) expansion of the heat kernel
\cite{Hawking:1976ja}.
The integral over the heat kernel ``time'' is cut off at $s=1/C^2$,
leading to the usual $C^4$, $C^2$ and $\log C$ divergences.
We have already seen that these divergences are present in our
calculation but get cancelled when the metric is put on shell.
Going on shell at fixed $C$ would not improve the situation
regarding the divergences.
However, if we take $C^2\sim R N^2$ and use the equations of motion at fixed $N$,
$C$ is seen to have a finite limit when $N\to\infty$ \cite{Becker:2020mjl}.
Since $C$ never goes to infinity, also the apparently divergent expressions
$C^4$, $C^2$ and $\log C$ are actually finite.
This shows that the finiteness of the quantum-corrected $m^2$ and $\lambda$
depends crucially on taking $N$ and not $C$ as the basic definition of cutoff.
One may add that just counting the number of field modes,
unlike choosing a maximal momentum,
is independent of the choice of units and is therefore a logically cleaner definition.

For the broken phase, one may contrast our calculation with \cite{Asorey:2012xq},
where the energy-momenum tensor of the broken phase had been studied.
They calculate separately the divergent and the finite part,
in view of keeping only the first as the physical, renormalized part,
while in our approach everything is rendered automatically finite
by the act of going on shell.
\smallskip

The cancellation of divergences that we observe for the scalar effective potential
is reminiscent of the very old notion of gravity as a universal regulator
\cite{Arnowitt:1960zz,DeWitt:1964yh,Isham:1970aw,Thiemann:1997rt}.
Looking for historical precedents, there is also some similarity between
the approach to the cosmological constant problem adopted here and some old ideas
of Adler \cite{Adler:1989rj} and Taylor and Veneziano \cite{Taylor:1989tm,Taylor:1989ua}.
The similarity lies in the presence, in the EA,
of nonlocal functions of the spacetime volume.
In particular, Taylor and Veneziano consider the consequences
of a term of the form (in our notation)
\be
\beta\Lambda^2 V_4\log(V_4/C^4)
\ee
where $\beta$ is some numerical constant.
This dependence on the volume is quite similar to the quantum term in (\ref{gammaN}),
that in the limit $N\to\infty$ can be rewritten
\be
-\frac{C^4 V_4}{1536\pi^2}\left[\log(V_4\mu^4)+constant\right]\ .
\ee
In both cases we have a $V_4\log V_4$ structure,
with the difference that in our case the cutoff appears outside the log,
whereas in their case it appeared inside.

We conclude with some comments on future extensions.
Whereas the generalization of these results to gravitons has already been 
considered in \cite{Becker:2021pwo}, it is of obvious interest to
look also at spinor and vector fields.
The main shortcoming of the present calculations is that they have
been derived in Euclidean signature.
In order to be more confident that the conclusions apply to the physical world,
it will be necessary to rederive them in Lorentzian signature.
As for the interactions, we have limited ourselves to a simple one loop calculation.
It will be interesting to see if the cancellations hold also at higher orders.

\section*{Acknowledgments}
We would like to thank D. Benedetti, V. Branchina, D. Ghilencea, S. P\"ogel and especially M. Reuter for useful discussions and comments.

\break

\appendix

%%%%%%%%%%%%%%%%%%%%%%
\section{Comparison with heat kernel evaluation}
%%%%%%%%%%%%%%%%%%%%%%

\subsection{The divergences}

Evaluating the EA with the Schwinger-De Witt method
and a UV momentum cutoff $\cut$, we obtain
\bea
\Gamma_{\cut}(R,\phi)&=&V_4\left[\frac{\Lambda_B}{8\pi G}
-\frac{1}{16\pi G}R+\frac12m^2\phi^2
+\frac{1}{24}\lambda\phi^4+\frac12\xi\phi^2 R\right]
\nonumber\\
&&-\frac{12}{R^2}
\bigg\{\frac{\cut^4}{2}
-\cut^2\left(m^2+\frac12\lambda\phi^2+R\left(\xi-\frac16\right)\right)
\nonumber\\
&&
+
\log\left(\cut^2/\mu^2\right)
\bigg[\frac{29}{2160}R^2
+\frac12\xi\left(\xi-\frac13\right)R^2
+\left(\xi-\frac16\right)\left(m^2+\frac12\lambda \phi^2\right)R
\nonumber\\
&&
+\frac18\lambda^2\phi^4
+\frac12\lambda m^2\phi^2
+\frac12m^4\bigg]\bigg\}+\ldots
\ ,
\eea
where the ellipses stand for subleading, finite terms.
The normal procedure is to treat $R$ and $\phi$ as externally fixed parameters,
to absorb the divergences in renormalized
parameters $\Lambda$, $G$, $m^2$, $\xi$, $\lambda$, 
and only then solve the resulting equations.
For the sake of comparison with the procedure in the text,
where we solve the equations at finite $N$ and then take the limit,
here we shall look at the solutions of the equations
for finite $C$ and then consider the limit.

It is still the case that $\phi=0$ is a solution of the $\phi$ equation of motion.
For simplicity we shall limit ourselves here to the symmetric phase.
The equation of motion of the metric,
evaluated at $\phi=0$, has the solution
\be
R=6\frac{8\pi\Lambda+2G m^2\cut^2-G\cut^4-G m^2\log\left(\cut^2/\mu^2\right)}{12\pi+(1-6\xi)G\cut^2+G m^2(6\xi-1)\log\left(\cut^2/\mu^2\right)}
\ee
If we turn off the gravitational interaction setting $G=0$,
this reduces to the classical solution $R=4\Lambda$.
For $\xi\not=1/6$, the leading behavior of this solution for large $\cut$ is
\be
\frac{\cut^2}{\xi-1/6}+O(\log\cut)\ .
\ee
This is the usual statement that curvature increases,
and the universe becomes smaller, as we increase the cutoff.
In the conformal case $m^2=0$ and $\xi=1/6$ the solution is instead
\be
R=4\Lambda-\frac{1}{2\pi}G\cut^4\ .
\ee

The effective (quantum corrected) mass and nonminimal coupling can be read off from
 \be
\frac{1}{V_4}\frac{\partial^2\Gamma_{\cut}}{\partial\phi^2}
\bigg|_{\phi=0}
=m^2+\xi R
+\frac{\lambda}{32\pi^2}\left[\cut^2
-\left(m^2+R\left(\xi-\frac16\right)\right)
\log\left(\cut^2/\mu^2\right)
\right]
\ .
\label{divcut}
\ee
The mass has both a quadratic and logarithmic divergence,
whereas the nonminimal coupling has only a logarithmic divergence:
\bea
m_{eff}^2&=&\frac{1}{V_4}
\frac{\partial^2\Gamma_C}{\partial\phi^2}
\bigg|_{\phi=0,R=0}
=m^2
+\frac{\lambda}{32\pi^2}\left[\cut^2
-m^2\log\left(\cut^2/\mu^2\right)
\right]+\ldots ,
\label{quadratic1}
\\
\xi_{eff}&=&\frac{1}{V_4}
\frac{\partial^3\Gamma_C}{\partial R\partial\phi^2}
\bigg|_{\phi=0,R=0}
=\xi
-\frac{\lambda}{32\pi^2}
\left(\xi-\frac16\right)
\log\left(\cut^2/\mu^2\right)+\ldots\ .
\label{quadratic2}
\eea
%Interpreting the effective couplings as renormalized couplings at scale $\mu$
%one can read off the standard beta functions
%\bea
%\mu\frac{\partial m^2}{\partial\mu}&=&\frac{\lambda}{16\pi^2}m^2
%\\
%\mu\frac{\partial \xi}{\partial\mu}&=&\frac{\lambda}{16\pi^2}\left(\xi-\frac16\right)
%\eea
The effective quartic coupling is
\be
\lambda_{eff}=\frac{1}{V_4}\frac{\partial^4\Gamma_C}{\partial\phi^4}
\bigg|_{\phi=0}
=\lambda
-\frac{3\lambda^2}{32\pi^2}
\log\left(\cut^2/\mu^2\right)+\ldots
\ .
\label{quartic}
\ee
%whence one gets the beta function:
%\be
%\mu\frac{\partial \lambda}{\partial\mu}=\frac{3\lambda^2}{16\pi^2}
%\ee

It is clear that in this approach, putting the background metric on shell
does not help at all. If anything, it makes the EA more divergenct.

\subsection{Recovering finiteness}

We shall now see how we can recover the main results of this paper
from this starting point.
We use the relation (\ref{ddcut}), that,
in view of taking the limit $N\to\infty$, we can simplify to
\be
C^2\sim\frac{R}{12}N^2\ .
\label{cutt}
\ee
We have seen that using the equation of motion
for $R$ including a finite number $N$ of modes,
$R$ decreases as $K_2/N^2$.
Then the key step is the observation that, using this relation,
the dimensionful cutoff (\ref{cutt}) has a finite $N\to\infty$ limit
\be
C^2\to \frac{K_2}{12}\ .
\label{Cfin}
\ee
In other words, if we take $N$ rather than $C$ 
to be the primary definition of cutoff,
and we use the equation of motion of $R$ before taking the limit,
we see that $C$ never goes to infinity.
Then, using (\ref{Cfin}) in (\ref{quadratic1}) and (\ref{quartic})
we obtain finite results
\bea
m_{eff}^2&=&m^2
+\frac{\lambda}{32\pi^2}\left[\frac{K_2}{12}
+m^2\log\left(\frac{12\mu^2}{K_2}\right)
\right]\ ,
%\\
%\xi_{eff}&=&\xi
%+\frac{\lambda}{32\pi^2}
%\left(\xi-\frac16\right)
%\log\left(\frac{12\mu^2}{K_2}\right)\ .
\\
\lambda_{eff}&=&\lambda
+\frac{3\lambda^2}{32\pi^2}
\log\left(\frac{12\mu^2}{K_2}\right)
\ .
\eea
If we choose the renormalization scale $\mu=m$,
the first of these is identical to (\ref{finitemass0}),
whereas the second differs from (\ref{finitel0}),
by a finite additive constant.

%%%%%%%%%%%%%%
\section{Some special functions}
%%%%%%%%%%%%%%

The harmonic number $H(n)$, for integer $n$, is defined as
\be
H(n) = \sum_{\ell= 1}^n \frac{1}{\ell}
\ee
It satisfies
\be
H(n+1)=H(n)+\frac{1}{n+1}\ .
\ee
Iterating this relation $m$ times we find
\be
 \sum_{\ell= 1}^m \frac{1}{\ell+n}=H(m+n)-H(n)\ .
\label{rel1}
\ee
Taking the limit $n\to\infty$ we find that
\be
\lim_{n\to\infty}(H(m+n)-H(n))=0\ .
\ee

One can extend the definition of $H(x)$ to the real domain in such a way that
these relations still hold for real arguments.
For $x\to\infty$, 
\be
H(x)\approx\log(x)+\gamma+\frac{1}{2x}-\frac{1}{12x^2}+\frac{1}{120x^4}+\ldots
\ee
where $\gamma\approx 0.5572$ is the Euler-Mascheroni constant, and for $a\to\infty$
\be
H(x+a)-H(a)\approx \frac{x}{a}-\frac{x(x+1)}{2a^2}+\frac{x(1+x)(1+2x)}{6a^3}+\ldots
\ee

We can use these properties to write an explicit expression for the function $\ess(N,z)$.
%In order to better understand the cancellations of the logarithmic divergences appearing in the harmonic numbers, we will analyze how they appear in the evaluation of the function $\mathbf{S} (N,z)$. 
Since in $\mathbf{S} (N,z)$ we have a second order polynomial in the denominator, we can find the root of that polynomial:
\be
\ell(\ell+3)+z = \left(\ell+x_+\right)\left(\ell+x_-\right)
\ee
where
\be
x_\pm=\frac{3\pm\sqrt{9-4z}}{2}\ .
\ee
Hence, by fraction decomposition
\be
\frac{1}{\ell(\ell+3)+z} = \frac{1}{\sqrt{9-4z}}\left(\frac{1}{\ell+x_-}-\frac{1}{\ell+x_+}\right)
\ee
Then, summing and using (\ref{rel1}) we obtain
\be
\sum_{N=1}^N\frac{1}{\ell(\ell+3)+z} 
= \frac{1}{\sqrt{9-4z}}\left(H(N+x_-)-H(x_-)-H(N+x_+)+H(x_+)\right)\ .
\ee
One can deal in a similar way with the sums that have $\ell$, $\ell^2$ or $\ell^3$
in the numerator, and one finds
\be
\ess(N,z)=\frac16\left\{N^2+4N+(2-z)\left[H(N+x_-)-H(x_-)-H(N+x_+)+H(x_+)\right]\right\}
\ .
\label{essexp}
\ee
When the metric is put on shell, $z$ diverges quadratically with $N$,
because it contain inverse curvature.
We are thus led to evaluate the function $\ess$ for $z=yN^2$,
where $y=\tfrac{12m^2}{K_2}$ is independent of $N$.
We obtain
\be
\ess\left(N,yN^2\right)
=\frac16\left[1-y\log\left(1+\frac{1}{y}\right)\right]N^2
+\frac{2}{3(1+y)}N
+\frac{10+57y}{36(1+y)^2}+\frac13\log\left(1+\frac{1}{y}\right)
+O(1/N)
\label{Sexp}
\ee
It is remarkable that although each of the harmonic numbers
diverges logarithmically with $N$, in the sum these divergences cancel exactly.
In fact, when we replace $z=y N^2$ (with $y>0$) the square bracket 
in (\ref{essexp}) reduces just to
\be
\log\left(1+\frac{1}{y}\right)\ .
\ee
In a similar way one can evaluate the derivative
\footnote{In fact using $\tfrac{\partial}{\partial z}=\tfrac{1}{N^2}\tfrac{\partial}{\partial y}$.
this could be obtained directly from (\ref{Sexp}).This is not an obvious statement, since $S$ is divergent.}
\be
\ess'(N,y N^2)\sim \frac16
\left[\frac{1}{1+y}+\log\left(1+\frac{1}{y}\right)\right]+O(1/N)\ .
\label{essprime}
\ee

%We will show that the sum over these two terms are both individually finite on-shell.
%\bea
%\sum_{\ell=1}^N\frac{1}{\ell+\frac{3-\sqrt{9-4z}}{2}} &=&\sum_{\ell=1+\frac{3-\sqrt{9-4z}}{2}}^{N+\frac{3-\sqrt{9-4z}}{2}}\frac{1}{k}   = -\sum_{\ell=1}^{\frac{3-\sqrt{9-4z}}{2}}\frac{1}{k}\; +\sum_{\ell=1}^{N+\frac{3-\sqrt{9-4z}}{2}}\frac{1}{k}  \\&=& -H\left(\frac{3-\sqrt{9-4z}}{2}\right)+H\left(\frac{3-\sqrt{9-4z}}{2}+N\right)
%\eea
%Now, on shell $z =y N^2$.
%\bea
% -H\left(\frac{3-\sqrt{9-4yN^2}}{2}\right)+H\left(\frac{3-\sqrt{9-4yN^2}}{2}+N\right)
%\eea
%Hence, expanding for large $N$, the leading order in $\log(N)$ will cancel the $N$ dependence and the different coefficients give rise to the $\log\left(1-\frac{1}{y}\right)$. As for the second term in A3, the same applies.

%%%%%%%%%%%%%%%%%%%%%%%%


\begin{thebibliography}{99}

\bibitem{Weinberg:1988cp}
S.~Weinberg,
``The Cosmological Constant Problem,''
Rev. Mod. Phys. \textbf{61} (1989), 1-23
%doi:10.1103/RevModPhys.61.1
%6256 citations counted in INSPIRE as of 17 Jul 2024

\bibitem{straumann}
N. Straumann, CERN lectures on Einstein's Impact on the Physics of the Twentieth Century (2005)
https://indico.cern.ch/event/425387/attachments/903020/1273882/lect.5.pdf

\bibitem{Zeldovich:1967gd}
Y.B.~Zeldovich,
``Cosmological Constant and Elementary Particles,''
JETP Lett. \textbf{6} (1967), 316

\bibitem{Akhmedov:2002ts}
E.K.~Akhmedov,
``Vacuum energy and relativistic invariance,''
arXiv:hep-th/0204048 [hep-th].

\bibitem{Ossola:2003ku}
G.~Ossola and A.~Sirlin,
``Considerations concerning the contributions of fundamental particles to the vacuum energy density,''
Eur. Phys. J. C \textbf{31} (2003), 165-175
%doi:10.1140/epjc/s2003-01337-7
[arXiv:hep-ph/0305050 [hep-ph]].

%\cite{Maggiore:2010wr}
\bibitem{Maggiore:2010wr}
M.~Maggiore,
``Zero-point quantum fluctuations and dark energy,''
Phys. Rev. D \textbf{83} (2011), 063514
%doi:10.1103/PhysRevD.83.063514
[arXiv:1004.1782 [astro-ph.CO]].
%89 citations counted in INSPIRE as of 17 Jul 2024

\bibitem{Asorey:2012xq}
M.~Asorey, P.~M.~Lavrov, B.~J.~Ribeiro and I.~L.~Shapiro,
``Vacuum stress-tensor in SSB theories,''
Phys. Rev. D \textbf{85} (2012), 104001
doi:10.1103/PhysRevD.85.104001
[arXiv:1202.4235 [hep-th]].

%\cite{DeWitt:1975ys}
\bibitem{DeWitt:1975ys}
B.~S.~DeWitt,
``Quantum Field Theory in Curved Space-Time,''
Phys. Rept. \textbf{19} (1975), 295-357
%doi:10.1016/0370-1573(75)90051-4
%1325 citations counted in INSPIRE as of 17 Jul 2024


\bibitem{Becker:2020mjl}
M.~Becker and M.~Reuter,
``Background Independent Field Quantization with Sequences of Gravity-Coupled Approximants,''
Phys. Rev. D \textbf{102} (2020) no.12, 125001
%doi:10.1103/PhysRevD.102.125001
[arXiv:2008.09430 [gr-qc]].

\bibitem{Becker:2021pwo}
M.~Becker and M.~Reuter,
``Background independent field quantization with sequences of gravity-coupled approximants. II. Metric fluctuations,''
Phys. Rev. D \textbf{104} (2021) no.12, 125008
%doi:10.1103/PhysRevD.104.125008
[arXiv:2109.09496 [hep-th]].

\bibitem{Banerjee:2023ztr}
R.~Banerjee, M.~Becker and R.~Ferrero,
``N-cutoff regularization for fields on hyperbolic space,''
Phys. Rev. D \textbf{109} (2024) no.2, 025008
%doi:10.1103/PhysRevD.109.025008
[arXiv:2302.03547 [hep-th]].

\bibitem{Weinberg:1987vp}
E.J.~Weinberg and A.q.~Wu,
``Understanding complex perturbative effective potentials,''
Phys. Rev. D \textbf{36} (1987), 2474

\bibitem{Benedetti:2014gja}
D.~Benedetti,
``Critical behavior in spherical and hyperbolic spaces,''
J. Stat. Mech. \textbf{1501} (2015), P01002
%doi:10.1088/1742-5468/2015/01/P01002
[arXiv:1403.6712 [cond-mat.stat-mech]].


\bibitem{Madore:1991bw}
J.~Madore,
``The Fuzzy sphere,''
Class. Quant. Grav. \textbf{9} (1992), 69-88
%doi:10.1088/0264-9381/9/1/0084

\bibitem{Fiore:2017ude}
G.~Fiore and F.~Pisacane,
``Fuzzy circle and new fuzzy sphere through confining potentials and energy cutoffs,''
J. Geom. Phys. \textbf{132} (2018), 423-451
%doi:10.1016/j.geomphys.2018.07.001
[arXiv:1709.04807 [math-ph]].


\bibitem{Pagani:2019vfm}
C.~Pagani and M.~Reuter,
``Background Independent Quantum Field Theory and Gravitating Vacuum Fluctuations,''
Annals Phys. \textbf{411} (2019), 167972
%doi:10.1016/j.aop.2019.167972
[arXiv:1906.02507 [gr-qc]].
%24 citations counted in INSPIRE as of 16 Apr 2024

%\cite{Ferrero:2022hor}
\bibitem{Ferrero:2022hor}
R.~Ferrero and M.~Reuter,
``The spectral geometry of de Sitter space in asymptotic safety,''
JHEP \textbf{08} (2022), 040
%doi:10.1007/JHEP08(2022)040
[arXiv:2203.08003 [hep-th]].
%10 citations counted in INSPIRE as of 18 Apr 2024

%\cite{Hawking:1976ja}
\bibitem{Hawking:1976ja}
S.W.~Hawking,
``Zeta Function Regularization of Path Integrals in Curved Space-Time,''
Commun. Math. Phys. \textbf{55} (1977), 133
%doi:10.1007/BF01626516

\bibitem{Arnowitt:1960zz}
R.~Arnowitt, S.~Deser and C.W.~Misner,
``Finite Self-Energy of Classical Point Particles,''
Phys. Rev. Lett. \textbf{4} (1960), 375-377
%doi:10.1103/PhysRevLett.4.375

\bibitem{DeWitt:1964yh}
B.~DeWitt,
``Gravity: A Universal regulator?,''
Phys. Rev. Lett. \textbf{13} (1964), 114-118
%doi:10.1103/PhysRevLett.13.114

\bibitem{Isham:1970aw}
C.J.~Isham, A.~Salam and J.A.~Strathdee,
``Infinity suppression in gravity modified quantum electrodynamics,''
Phys. Rev. D \textbf{3} (1971), 1805-1817
%doi:10.1103/PhysRevD.3.1805

\bibitem{Thiemann:1997rt}
T.~Thiemann,
``QSD 5: Quantum gravity as the natural regulator of matter quantum field theories,''
Class. Quant. Grav. \textbf{15} (1998), 1281-1314
%doi:10.1088/0264-9381/15/5/012
[arXiv:gr-qc/9705019 [gr-qc]].
%394 citations counted in INSPIRE as of 16 Apr 2024

\bibitem{Adler:1989rj}
S.L.~Adler,
``Effective Action Model for the Vanishing of the Cosmological Constant,''
Phys. Rev. Lett. \textbf{62} (1989), 373-375
%doi:10.1103/PhysRevLett.62.373

\bibitem{Taylor:1989tm}
T.R.~Taylor and G.~Veneziano,
``Quenching the Cosmological Constant,''
Phys. Lett. B \textbf{228} (1989), 311-316
%doi:10.1016/0370-2693(89)91551-7

\bibitem{Taylor:1989ua}
T.R.~Taylor and G.~Veneziano,
``Quantum Gravity at Large Distances and the Cosmological Constant,''
Nucl. Phys. B \textbf{345} (1990), 210-230
%doi:10.1016/0550-3213(90)90615-K


\end{thebibliography}
\end{document}